\documentclass[twocolumn]{aastex631}

\definecolor{cadmiumgreen}{rgb}{0.0, 0.42, 0.24}

\usepackage{comment}
\usepackage{float}
\usepackage{amsmath}

\begin{document}

\title{Black Hole Supernovae, their 
Equation of State Dependence and Ejecta Composition}

\correspondingauthor{Oliver Eggenberger Andersen}
\email{oliver.e.andersen@astro.su.se}

\author[0000-0002-9660-7952]{Oliver Eggenberger Andersen}
\affil{The Oskar Klein Centre, Department of Astronomy,
Stockholm University, AlbaNova, SE-106 91 
Stockholm, Sweden}

\author[0000-0002-8228-796X]{Evan O'Connor}
\affil{The Oskar Klein Centre, Department of Astronomy,
Stockholm University, AlbaNova, SE-106 91 
Stockholm, Sweden}

\author[0000-0002-4747-8453]{Haakon Andresen}
\affil{The Oskar Klein Centre, Department of Astronomy,
Stockholm University, AlbaNova, SE-106 91 
Stockholm, Sweden}

\author[0000-0003-0849-7691]{André da Silva Schneider}
\affil{Departamento de Física, Universidade Federal de Santa Catarina, Florianópolis, SC 88040-900, Brazil}

\author[0000-0002-5080-5996]{Sean M. Couch}
\affil{Department of Physics and Astronomy, Michigan State University, East Lansing, MI 48824, USA}
\affil{Department of Computational Mathematics, Science, and Engineering, Michigan State University, East Lansing, MI 48824, USA}
\affil{Facility for Rare Isotope Beams, Michigan State University, East Lansing, MI 48824, USA}

\begin{abstract}

Recent literature on core-collapse supernovae suggests that a black hole (BH) can form within $\sim 1$\,s of shock revival, while still culminating in a successful supernova. We refer to these as black hole supernovae, as they are distinct from other BH formation channels in both timescale and impact on the explosion. We simulate these events self-consistently from core-collapse until 20--50\,days after collapse using three axisymmetric models of a 60 M$_\odot$ zero-age main sequence progenitor star and investigate how the composition of the ejecta is impacted by the BH formation. We employ Skyrme-type equations of state (EOSs) and vary the uncertain nucleonic effective mass, which affects the pressure inside the proto-neutron star through the thermal part of the EOS. This results in different BH formation times and explosion energies at BH formation, yielding final explosion energies between $0.06\text{--}0.72\times 10^{51}$\,erg with $21.8\text{--}23.3\,M_\odot$ of ejecta, of which $0\text{--}0.018\,M_\odot$ is $^{56}$Ni. Compared to expectations from 1D simulations, we find a more nuanced EOS dependence of the explosion dynamics, the mass of the BH remnant, and the elemental composition of the ejecta. We investigate why the explosions survive despite the massive overburden and link the shape of the diagnostic energy curve and character of the ejecta evolution to the progenitor structure.\\

\end{abstract}
 
\section{Introduction} 
\label{sec:intro}
Core-collapse supernovae (CCSNe) play an important role in the Universe.  The supernova ejecta carries the yields of nucleosynthesis processes occuring throughout the star's life, including at the end, that contribute to the enrichment of the Universe. Furthermore, neutron stars (NSs) and stellar-mass black holes (BHs) are born in CCSNe. Data analysis from X-ray binary observations \citep{oezel:16,lattimer:21} and gravitational wave interferometers are beginning to map out the population of NSs and BHs \citep{abbott:23co_distribution}. The mass distributions of NSs and BHs will be increasingly resolved with more observations and with future detectors \citep{shao:22, abbott:23_O3-2}. To predict and understand the shape of these distributions, it is crucial that we take into account and understand all NS and BH formation channels. In the conventional theory of CCSNe, the outcome of core collapse is dichotomous: Either the shock wave unbinds the envelope of the star and a NS is left behind, or the stalled shock is not revived and continued accretion onto the proto-neutron star (PNS) results in BH formation, called failed supernovae \citep{oconnor:11}. Failed supernovae are often associated with either dim or no electromagnetic signals \citep{byrne:22}, and several studies have searched for and found ``disappearing stars" \citep{kochanek:08, reynolds:15, adams:17, basinger:21, neustadt:21, de:24}. However, there is still no conclusive evidence connecting these candidates to failed supernovae \citep{byrne:22}.

Recent advancements in the field, however, have revealed a highly diverse landscape of CCSN outcomes. BHs can form via several paths, all of which need to be accounted for in compact object population synthesis calculations \citep{mapelli:21, belczynski:22}. In events where the shock is not revived, the outer layers of the star can be expelled as a hydrodynamical response to the neutrino emission in the early post-bounce phase \citep{nadezhin:80,piro:13c, lovegrove:13}, which can result in a transient and a BH mass that differs from the progenitor mass \citep{lovegrove:13, fernandez:18, ivanov:21, schneider:23}. 

BHs can form in various ways when the shock is revived. For instance, thermal pressure can temporarily hold a NS above the maximum mass that the cold equation of state (EOS) can support, leading to its collapse into a BH once it cools \citep{baumgarte:96, oconnor:11proc}. Alternatively, BH formation can occur when initially outgoing matter falls below the escape velocity \citep[e.g.][]{colgate:71, chevalier:89b, fryer:99, zhang:08, ertl:20}, eventually accreting and pushing the NS above the maximum mass that the EOS can support. These events, known as fallback supernovae, occur over tens of seconds to hours after shock revival depending on the fallback mechanism \citep{wong:14}. In fallback supernovae, the long timescales between explosion and BH formation limit the impact on the explosion.

A new BH formation channel that is occurring on much shorter timescales than fallback supernovae and significantly affecting explosion dynamics is beginning to be unveiled. In three-dimensional (3D) simulations, \citet{chan:18}, \citet{chan:20}, and \citet{burrows:23} demonstrate the possibility of having a successful explosion, while continued accretion of the progenitor core induces BH formation within 0.55--1.5\,s after the shock wave is launched. Despite the quick BH formation, the explosions remain successful and their final explosion energies are comparable to that of a garden-variety supernova. Multi-dimensional dynamics appear crucial in precipitating these BH-forming supernovae. They enable the explosion to develop along specific solid angles, while the bulk of accretion occurs from other directions, creating conditions for a non-spherical explosion morphology.

Other studies hint towards the viability of these SNe with rapid BH formation as well, where shock revival precedes BH formation in two-dimensional (2D) simulations \citep{pan:18, rahman:22, sykes:23, sykes:24} and in 3D \citep{kuroda:18, walk:20, pan:21, powell:21}. A high compactness ($\xi_{2.5} = 0.86$; see Equation~\ref{eq:compactness}) 85\,M$_\odot$ progenitor is simulated by \cite{powell:21} in 3D and by \cite{sykes:23} in 2D. In both studies, BH formation occurs hundreds of milliseconds after shock revival. The times of shock revival and the times of BH formation are very similar between the two studies (for the same EOS). As the binding energy of the layers above the shock is higher than the diagnostic explosion energy at BH formation, the final fates of the explosions in Powell et al. and Sykes et al. are uncertain. \cite{powell:21} discuss various outcome scenarios involving possible mass ejection. They show that the BH formation time increases in order with the LS220, SFHo, and SFHx EOSs, and point to the thermal dependence of the EOSs as the underlying cause. \cite{rahman:22} simulate high compactness progenitors ($\xi_{2.5}$ between 0.77 and 0.89) using the SFHo EOS, and witness BH formation within hundreds of milliseconds after shock revival. Their diagnostic explosion energy drops to zero within a few seconds. Still, a shock or sonic wave moves outwards carrying $4\text{--}7\times10^{49}$\,erg, and they estimate mass losses between $0.07\text{--}3.5$\,M$_\odot$. Recently, \citet{sykes:24} conducted 2D simulations of eight compact ($\xi_{2.5} > 0.62$) zero-metallicity progenitors with masses ranging from 60\,M$_\odot$ to 95\,M$_\odot$ using the LS220 EOS. In their simulations, BH formation occurs between 0.12\,s and 2.1\,s after shock revival, with all explosions surviving and resulting in final explosion energies between $\sim0.4\times10^{51}$\,erg and $\sim2.5\times10^{51}$\,erg.\\

We continue to explore this novel BH formation channel. It is distinct from fallback supernovae in that it is direct accretion of the collapsing progenitor that causes BH formation, and not ejected material that reverses and accretes at later times. In fallback supernovae, BH formation has a small impact on the overall explosion dynamics, whereas the BH formation channel explored here has a large impact on the explosion properties. To place this BH formation channel in a category of its own, between failed supernovae and fallback supernovae, we will refer to them as black hole supernovae (BHSNe).

When considering the viability and prevalence of BHSNe, part of the answer lies in understanding which progenitors explode and which do not---a quest with a long history that is still ongoing \citep{fryer:99, oconnor:11, ugliano:12, pejcha:15, perego:15, nakamura:15, ertl:16, sukhbold:16, mueller:16c, curtis:19, ebinger:19, couch:20, boccioli:21, boccioli:23}.  However, more light is being shed on this question with each state-of-the-art simulation and explodability prediction model, along with insights gained from theoretical work on the imbalance of forces that often results in explosions when the silicon-oxygen interface accretes \citep{gogilashivili:24, boccioli:24}. 
Recent state-of-the-art simulations indicate that explosions do not only occur for low-compactness progenitors, but may take place across the entire compactness range, with a lower tendency to explode in the middle of the range \citep{burrows:19, burrows:23}. High compactness progenitors are particularly interesting for BHSNe due to their high accretion rates, which may fuel an energetic explosion and simultaneously push the proto-neutron star (PNS) above the maximum mass that the combined thermal and non-thermal parts of the EOS can support.\\

Several studies have shown that the thermal part of the EOS is a significant uncertainty in the outcome of SNe and their multi-messenger signals \citep{schneider:19, schneider:20, yasin:20, eggenbergerandersen:21}. Specifically, \citet{schneider:19} constructed 97 Skyrme type \citep{skyrme:62} finite-temperature EOSs to systematically explore the impact of the parameters that characterize an EOS by performing a Taylor expansion of the energy functional about nuclear saturation density. For the progenitor used ($\xi_{2.5} = 0.2785$), \citet{schneider:19} find that the uncertainty range for the effective mass of nucleons ($m^{\star}$), which determines the thermal dependence of their EOSs, has a dominant impact on CCSNe compared to the non-thermal parameters and their uncertainties. Since the PNS temperature generally increases with compactness, the effective mass and thermal part have an even stronger effect in higher compactness models, as demonstrated by \citet{schneider:20} in their exploration of the EOS impact on BH formation in spherically symmetric simulations.\\

In this study, we use a high-compactness 60\,M$_\odot$ progenitor and investigate how the thermal part of the EOS affects the outcome of BHSNe. We self-consistently simulate the collapse and shock revival phase, as well as the evolution after BH-formation until days after shock breakout. We demonstrate how the explosion dynamics, mass of the BH remnant, final explosion energy, and composition of the supernova ejecta all depend on the thermal part of the EOS. \\

The remainder of the paper is structured as follows: In Section~\ref{sec:methods}, we detail the methodologies used before and after BH formation, the EOSs employed, the calculations for nucleosynthesis, and features of the progenitor. In Sections~\ref{sec:prior} and~\ref{sec:afterBH}, we present the results before and after BH formation, respectively. We analyze the evolution of the ejecta composition in Section~\ref{sec:ejecta}. In Section~\ref{sec:discussion}, we discuss how the explosions survive through the massive overburden and how our results may transfer to 3D. We summarize and conclude in Section~\ref{sec:conclusions}.

\vspace{1cm}
\section{Methods}
\label{sec:methods}

We utilize the FLASH (v.4) simulation framework \citep{fryxell:00, dubey:09}, modified for simulating CCSNe \citep{couch:13a, couch:14a, oconnor:18, oconnor:18c}. We relay the methods in two parts: (1) The collapse phase until BH formation which captures the evolution until $\sim1.5$\,s post bounce. (2) The post-BH formation phase, where we mask the BH central region to track the evolution of the explosion without an energy source from the PNS. This phase captures the evolution between at least $\sim 1.5$\,s and $\sim22$ days, illustrating how the shock disrupts the star.

\subsection{Evolution before black hole formation}

We evolve the collapse phase in one dimension (1D) until 20\,ms after bounce, from which point we continue the simulations in 2D under the assumption of axisymmetry in cylindrical geometry. The domain extends $\pm 8\times10^{11}$\,cm along the cylindrical axis and $8\times10^{11}$\,cm along the radial axis and is placed onto a grid of $20\times10$ blocks. Each block has 16x16 zones. 18 levels of adaptive mesh refinement are allowed, yielding a finest grid spacing of 281\,m. Mesh refinements occur based on density and pressure gradients, and above 90\,km we limit the resolution so that an effective resolution of $\sim0.6^\circ$ is maintained.
We utilize Newtonian hydrodynamics with an effective general relativistic gravitational potential \citep[‘Case A’ in][]{marek:06}. This approach has been demonstrated to effectively capture the general relativistic effects in core-collapse supernovae \citep{marek:06,mueller:12a,oconnor:18}. There is good agreement in BH formation times in 1D when comparing an effective potential and a full general relativistic treatment \citep{schneider:20}. We propagate the time evolution using a second order Runge-Kutta method \citep{couch:21}. Reconstruction is performed with the WENO-Z method \citep{borges:08} and fluxes between zones are calculated with the HLLC Riemann solver \citep{toro:94}. For zones near shocks, the HLLE Riemann solver \citep{HLLE:88} is used. For neutrino transport we evolve the first two moments of the neutrino distribution function with an analytic M1 closure \citep{shibata:11,oconnor:18}. 12 logarithmically spaced energy groups between 1\,MeV and $\sim315$\,MeV are used for three species of neutrinos: electron neutrinos ($\nu_e$), electron antineutrinos ($\bar{\nu}_e$), and the composite group of heavy lepton neutrinos and their antineutrinos ($\nu_x$). Neutrino opacities are calculated using \texttt{NuLib} \citep{oconnor:15a} and are the baseline set from \cite{oconnor:18c}, with the inclusion of effective mean field and virial corrections \citep{horowitz:2017} as well as inelastic scattering of neutrinos on electrons following \citet{bruenn:85}.

When performing initial 1D tests with a wide range of time-steps and several time-evolution solvers in the neutrino transport, we found that the time of BH formation can shift by up to 100\,ms depending on the setup. Decreasing the time-step led to faster BH formation across all solvers. In terms of BH formation times, we found that the two-step forward Euler method is more consistent than a standard second order Runge-Kutta method, which we link to the velocity-dependent treatment of the fluxes in the asymptotic limit. Therefore, we use the two-step forward Euler scheme for the 2D simulations, with a time-step size such that the corresponding 1D simulation forms a BH within 30\,ms of the earliest BH formation time we observed across all the 1D simulations. It is unclear whether the effect observed in 1D persists in 2D or is washed out when an extra degree of freedom is included. Nevertheless, our 1D test simulations highlight an important, albeit small, source of uncertainty in numerical simulations.

\subsection{Evolution after black hole formation}
Similar to the approach of \citet{chan:18, rahman:22, sykes:23}, and more specifically \citet{schneider:23}, we excise the central region when the BH forms and continue the evolution until at least $\sim22$\,days. At BH formation we set the radius of this mask to 150\,km and increase its size by 100\,km\,s$^{-1}$, imposing a maximum radius of $2\times10^6$\,km. To verify that this mask size and its edge velocity are reasonably large, we run a model for 8\,s with a mask radius of 50\,km and expansion velocity of 50\,km\,s$^{-1}$. The explosion energy evolution is nearly indistinguishable from the production run in this paper. 

The material inside the mask is collected into a point mass, as is the material that subsequently accretes into the ``BH". We create a low-pressure region inside the mask by setting the density, temperature and internal energy to a small fraction of their values outside the mask. This allows the material to freely flow into the mask and is effectively an outflow boundary condition. We do not evolve the neutrino radiation after BH formation because, to a good approximation, there is no longer a source for neutrinos. Since \texttt{FLASH} is fundamentally a Newtonian hydrodynamics code, we estimate the gravitational mass ($M_\mathrm{grav}$) of the compact object at time $t$ by subtracting the mass-energy radiated away by neutrinos from the baryonic mass ($M_\mathrm{bary}$): 
\begin{equation}\label{eq:mgrav}
M_\mathrm{grav} = M_\mathrm{bary} - \frac{1}{c^2}\sum_\nu \int_0^t L_\nu(t^\prime)dt^\prime,
\end{equation}
where $L_\nu$ is the neutrino luminosity from species $\nu$. We get the point mass at the time of BH formation ($t_\mathrm{BH}$) by inserting $t=t_\mathrm{BH}$ into Equation~\ref{eq:mgrav}. 

When applying the excision mask at BH formation for each simulation, we also extend the domain by creating a larger grid filled with the progenitor data, then map the simulated 2D data onto this grid. At this time we extend the grid from $8\times10^{11}$\,cm to $1.024\times10^{14}$\,cm. We perform one more such mapping at $\sim700$\,s, extending the domain out to $8.192\times10^{14}$\,cm. We add a low-density, artificial 2.1\,M$_\odot$ atmosphere outside the surface of the star at $1.79\times10^{14}$\,cm. We set the temparature ($T$) and electron fraction ($Y_e$) in the atmosphere to the surface values and assign a density ($\rho$) of $1.2\times10^{-12}$g\,cm$^{-3}$ (close to the minimum of the EOS). We arbitrarily designate the material in the atmosphere as neutrons to avoid interference with other species when tracking the composition of the ejecta. With this atmosphere in place we follow the evolution until days after the shock breakout. At the time of the second mapping, at $\sim700$\,s, we change the reconstruction method from the spatially fifth order accurate WENO-Z to a second order accurate TVD \citep{harten:83} due to issues encountered at the sharp density interfaces in the ``Rayleigh-Taylor" bullets occurring behind the reverse shock \citep[see e.g][]{kifonidis:03, sandoval:21}. Although many such bullets are created in our simulations at late times, fewer emerge when using the more diffusive TVD reconstructor compared to when using WENO-Z. Switching to TVD significantly earlier than $700$\,s leads to a slight difference in the explosion energy. Although there is no significant effect on the total mass accretion evolution, fewer heavy elements (such as $^{56}$Ni) accrete into the BH. Switching after $700$\,s does not impact the composition or the explosion energy.

\subsection{Equation of State}
\label{sec:eos}

The hydrodynamic equations are closed using a hybrid EOS \citep{witt:21}. At low densities, we use the $Y_e$ based Helmholtz EOS \citep{timmes:00}, which bridges at a transition density to a Skyrme-type nuclear EOS for higher densities. We refer to \citet{schneider:19,schneider:20} and \citet{eggenbergerandersen:21} for details on the nuclear EOSs used and describe the most relevant aspects for this work here. 

We deploy the baseline EOS from \citet{schneider:19} in the \texttt{m0.75} model, where the effective mass of nucleons for symmetric matter at nuclear saturation density is 0.75, expressed as a fraction of the vacuum mass. Keeping all other parameters in the EOS fixed, we vary the effective mass by 2\,$\sigma$ above (model \texttt{m0.95}) and below the baseline value (model \texttt{m0.55}). The effective masses of neutrons and protons, $m_n^{\star}$ and $m_p^{\star}$, primarily determine the temperature-dependent part of the EOS via the kinetic energy per baryon term, $\epsilon_{\mathrm{kin}}$:

\begin{equation}
\begin{split}
\epsilon_{\mathrm{kin}}(n,y,T) &= \frac{\sqrt{2}~T^{5/2}}{\pi^2\hbar^3 n} \Big({m_n^{\star}}^{3/2}F_{3/2}(\eta_n) \\
&\quad \hspace{2cm} + {m_p^{\star}}^{3/2}F_{3/2}(\eta_p)\Big).
\end{split}
\label{eq:e_kin}
\end{equation}
Here, $F_{3/2}$ is a Fermi integral using the degeneracy parameters $\eta_n$  and $\eta_p$ \citep[see][]{schneider:19}, $n$ is the baryon number density, $T$ is the temperature, and $y$ is the proton fraction which enters into the expression for the effective mass:

\begin{equation}\label{eq:meff}
\frac{\hbar^2}{2m_t^{\star}} = \frac{\hbar^2}{2m_t} + \alpha_1 n_t+ \alpha_2 n_{-t}\, ,
\end{equation} 
where $\alpha_i$ are parameters in the Skyrme model and $m_t$ are the nucleon vacuum masses. If $t = n$ then $-t = p$ and vice versa. This is only one possible parametrization of the effective mass, which depends on the proton and neutron densities. More complex parametrizations exist \citep{Constantinou:15}.

\subsection{Determining Nucleosynthetic Yields}\label{sect:methods-nucleosynthesis}

\begin{figure}[b!]
    \centering
    \includegraphics[width=\columnwidth]{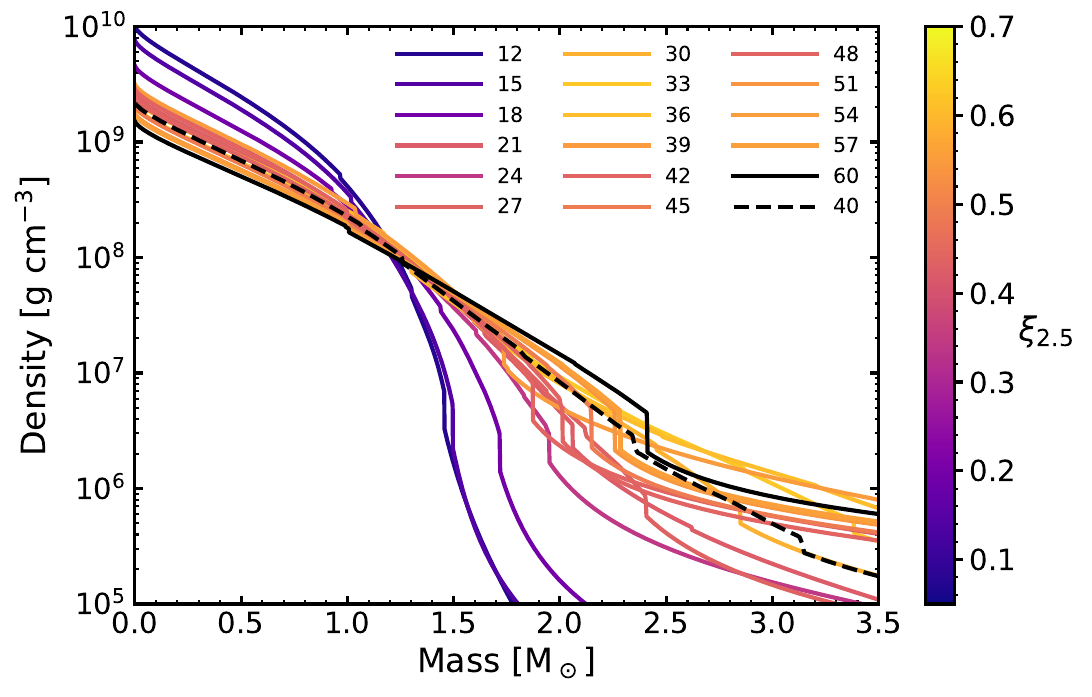}
    \caption{For several progenitors from \cite{sukhbold:18}, the density profiles are plotted and colored according to their compactness as defined by Equation~\ref{eq:compactness} using a 2.5\,M$_\odot$ mass coordinate. In black we highlight the $\xi_{2.5} = 0.63$, 60\,M$_\odot$ progenitor we  use in this study. We also include the $\xi_{2.5} = 0.54$, 40\,M$_\odot$ progenitor from \cite{sukhbold:16} (dashed black line) that results in a BHSN in \citet{burrows:23}.} 
    \label{fig:compactness}
\end{figure}

\begin{figure*}[]
    \centering
    \includegraphics[width=0.8\linewidth]{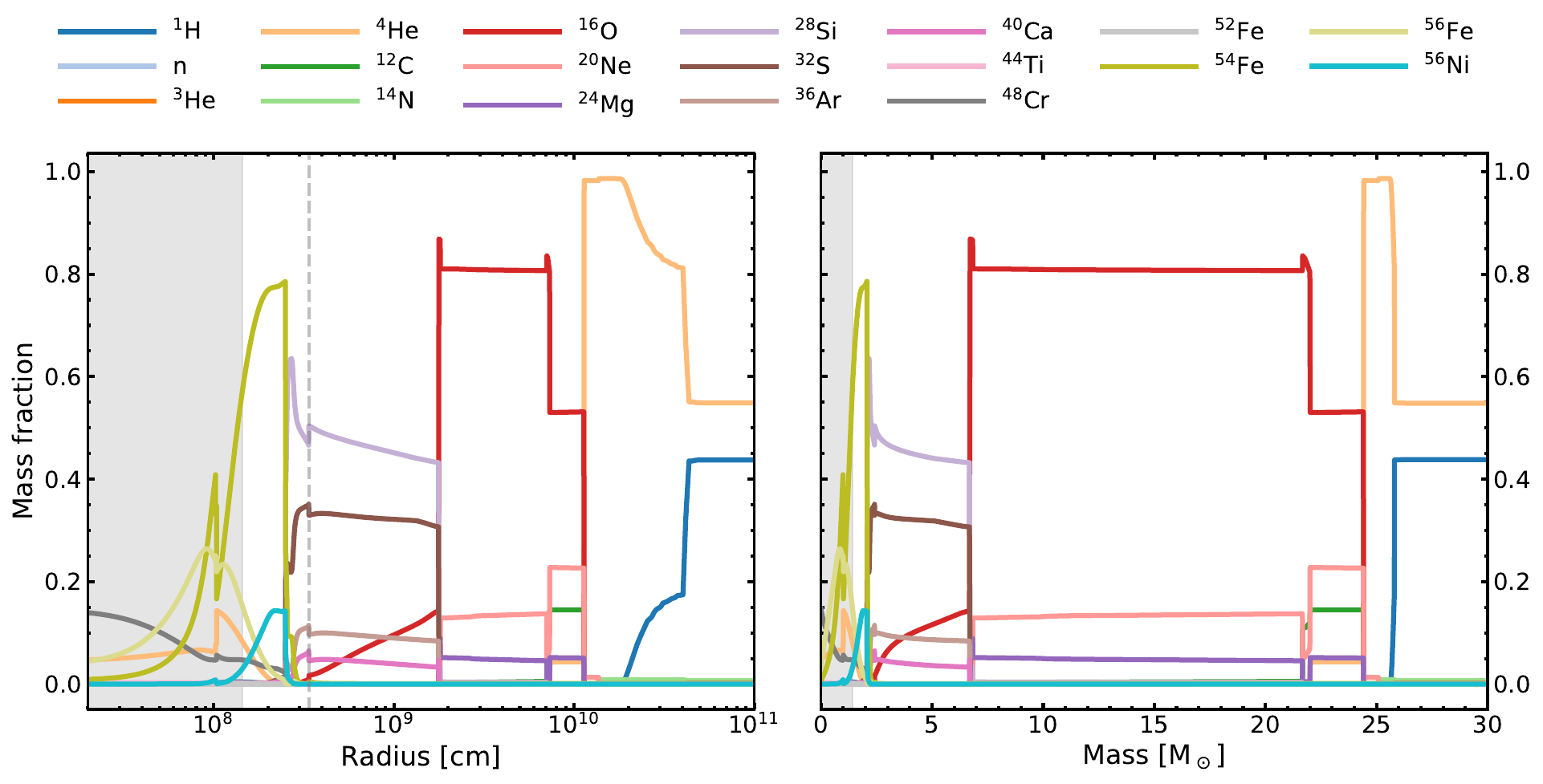}
    \caption{Mass fractions of the elements in the progenitor at the point of collapse, as a function of radius (left) and enclosed mass (right). Outside of $10^{11}$\,cm and 30\,M$_\odot$, where we limit these plots, the $^{4}$He and $^{1}$H mass fractions are approximately constant. The dashed vertical line at $\sim 3.4\times 10^{8}$\,cm ($\sim 2.4$\,M$_\odot$) in the silicon layer, corresponds to the sharp drop in the density profile in Figure~\ref{fig:compactness} and marks the transition from the oxygen burning layer to the silicon burning layer. The shaded areas indicate regions where more than 10\% of the mass consists of other species, as calculated in a quasi-statistical equilibrium network \citep{weaver:78, sukhbold:18}.}
    \label{fig:progenitor_composition}
\end{figure*}
At the time we transition from the core-collapse evolution to the post core-collapse evolution we populate the entire simulation domain with a detailed set of compositions that we then advect through the subsequent explosion phase. \cite{sieverding:23} show that an alternative method to using tracer particles is to generate the particle trajectories during post-processing by integrating the equations of motion, i.e. $d\vec{x}(t)/dt = \vec{v}$, backwards (or forwards) in time.  The backward integration method is ideal for systematically populating a domain at a given time since one can control the location where the yields are to be determined.  This is the method we utilize here.

In our case, we use backwards integration of the equations of motion to obtain a representative past history of each computational zone. To limit the number of total trajectories, we only have one trajectory for each 2$\times$2 computational zones and we assume the yields are then constant over the four zones. In total, each of our simulations use $\sim$50,000 trajectories.  Along each trajectory we record the matter density ($\rho$), temperature ($T$), electron fraction ($Y_e$), and the change in the electron fraction coming from the neutrino interactions ($dY_e^\nu$). The electron fraction evolves along the trajectory.  Some of this evolution is from the aforementioned $dY_e^\nu$. However, the electron fraction evolution along the trajectory can also suffer from hydrodynamical mixing of the fluid, generally smoothing out the electron fraction and preventing the most neutron- or proton-rich trajectories from surviving \citep{stockinger:20}. Recording $dY_e^\nu$ along the trajectory allows to us assess the impact of this. We stop the backwards integration when the trajectories reach a matter temperature of 6\,GK, or the beginning of the simulation.  We then process each trajectory through SkyNet \citep{lippuner:17b}.  We model our setup after the X-ray burst test case provided with SkyNet.  This utilizes a 686 species reaction network, with over 8000 reactions between the species.  For each trajectory we initialize the network with either NSE compositions (if the trajectory originates at a matter temperature of 6\,GK) or the initial composition from the progenitor (based on the initial radius).  For the NSE initial conditions we assume the $Y_e$ of the trajectory at 6\,GK while for the initial composition at the start of collapse we take the $Y_e$ from the initial conditions.  In both cases $Y_e$ is not evolved along the trajectory within SkyNet, this is a good approximation as most neutrino interactions will happen at temperatures well above the 6\,GK NSE transition temperature we choose here. The precision predictions of many isotopes \citep{froehlich:06,sandoval:21,harris:17} will require a higher transition temperature and a self-consistent coupling of neutrino interactions into the nuclear reaction network.  

Following the SkyNet calculation, we reduced the compositions to 20 species, plus 2 species to track the remaining mass. The two other species fall into two categories: ``fossil" elements, which are elements more massive than $^{44}$Ti present outside the silicon core in the progenitor star, and ``tracer" elements, which are elements outside of our standard set of 20 elements that are produced within SkyNet in the nucleosynthesis calculation.  The need of ``fossil" elements is due to the available set of initial compositions in the progenitor model that do not accurately reflect the actual distribution of these fossil elements. Our initial models have $\sim 0.05$\,M$_\odot$ of these fossil elements, which can be considered to have the solar metallicity distribution of elements more massive than $^{44}$Ti. These species are simply advected in our post core-collapse simulations.

\subsection{Progenitor}\label{sect:progenitor}

We use a 60\,M$_\odot$ zero-age main sequence solar metallicity progenitor from \cite{sukhbold:18} with reduced mass loss (from their $\dot{M}_\mathrm{N}/10$ models), resulting in a remaining mass of 47.31\,M$_\odot$ by the time the core begins collapsing. This progenitor has a compactness, $\xi_{2.5} = 0.63$ , and $\xi_{1.75} = 0.91$, with compactness defined as in \cite{oconnor:11},

\begin{equation}\label{eq:compactness}
    \xi_{M} = \frac{M/M_\odot}{R(M_{\mathrm{bary}} = M)/1000\,\mathrm{km}},
\end{equation}
and evaluated at the time of collapse. This is the highest 2.5-M$_\odot$ compactness progenitor in the compendium from \citet{sukhbold:18}, but it is worth noting that there are similarly compact progenitors in the $30\text{-}40$\,M$_\odot$ mass range. This can be seen in Figure~\ref{fig:compactness} (and in Figure~8 of \citealt{sukhbold:18}) where we plot density profiles from the $\dot{M}_\mathrm{N}/10$ compendium. We highlight the 60\,M$_\odot$ progenitor used in this study with a black line color. For comparison, which we return to later in Section~\ref{sec:discussion}, we also include a 40\,M$_\odot$, $\xi_{2.5} = 0.54$ , and $\xi_{1.75} = 0.87$ progenitor from \cite{sukhbold:16} (dashed black line). This 40\,M$_\odot$ progenitor was simulated in 3D by \citet{burrows:23} and resulted in a BHSN. For the 60\,M$_\odot$ progenitor, there is a sharp drop in density at $\sim 2.4$\,M$_\odot$, which is near the bottom of the silicon/sulfur layer and does not correspond to the silicon-oxygen interface (at $\sim6.7$\,M$_\odot$). Figure~\ref{fig:progenitor_composition} shows the initial composition profile of the progenitor in both radius coordinate (left) and enclosed mass coordinate (right). The dashed vertical line in the left panel marks the radial coordinate ($\sim 3.4\times 10^{8}$\,cm) that corresponds to the $\sim 2.4$\,M$_\odot$ mass coordinate. In the silicon/sulfur layer, which is bounded by the iron layer from below and the oxygen layer from above, there is yet a small mass fraction of oxygen. We interpret the sharp drop in the density profile at $\sim 2.4$\,M$_\odot$ as the transition between the region where oxygen is being burnt to silicon and the region where silicon is being burnt to iron.

\section{Results}

\begin{table}[ht!]
     \centering
     \begin{tabular}{cccccccc}
     \hline \hline
       \multicolumn{1}{c}{\textbf{$m^\star$}} & \textbf{$t_{\mathrm{sr}}$} & \textbf{$t_{\mathrm{BHF}}$} & \textbf{$E_{\mathrm{expl}}$} & \textbf{$M_{\mathrm{BH}}$}  & \textbf{$M_{\mathrm{ej}}$} & \textbf{$M_{\mathrm{Ni^{56}}}$}\\
        \multicolumn{1}{c}{[$m_n$]} & [s] & [s] &  [$10^{50}$\,erg] & [M$_\odot$] & [M$_\odot$] & [M$_\odot$]\\ \hline 
        \texttt{0.95} \vline & 0.292 & 0.93  & 0.57 & 25.48  &21.67 & 0.000\\ 
        \texttt{0.75} \vline & 0.317 & 1.36  & 7.23 & 24.02  &23.09 &  0.0157 \\
        \texttt{0.55} \vline & 0.346 & 1.45   & 6.68 & 24.30  &22.80 & 0.0182\\ \hline \hline
     \end{tabular}
     \caption{Overview of the simulations that vary in the effective mass parameter, $m^\star$. We report the approximate shock revival time, $t_{\mathrm{sr}}$, the time to BH formation ($t_{\mathrm{BHF}}$), the final explosion energy ($E_{\mathrm{expl}}$), the final BH mass ($M_{\mathrm{BH}}$), the ejected mass ($M_{\mathrm{ej}}$), and the $^{56}$Ni ejecta ($M_{\mathrm{Ni^{56}}}$). Times are relative to core bounce.}
     \label{table:simulations}
   \end{table}

\begin{figure}[ht!]
    \centering
    \includegraphics[width=\columnwidth]{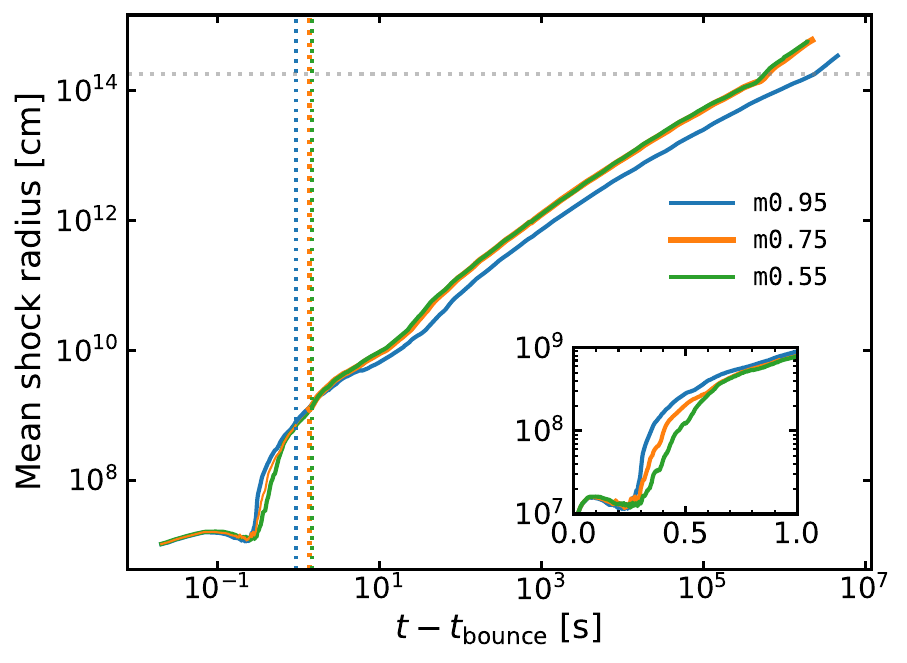}
    \caption{Mean shock radius as a function of time for the three models varying only in the effective mass parameter in the EOS. Increasing the effective mass thermally softens the EOS, increasing the explodability. The vertical dotted lines indicate the time of BH formation for each model. The horizontal line marks the surface of the star.} 
    \label{fig:shock_rad}
\end{figure}

\begin{figure}[ht!]
    \centering
    \includegraphics[width=\columnwidth]{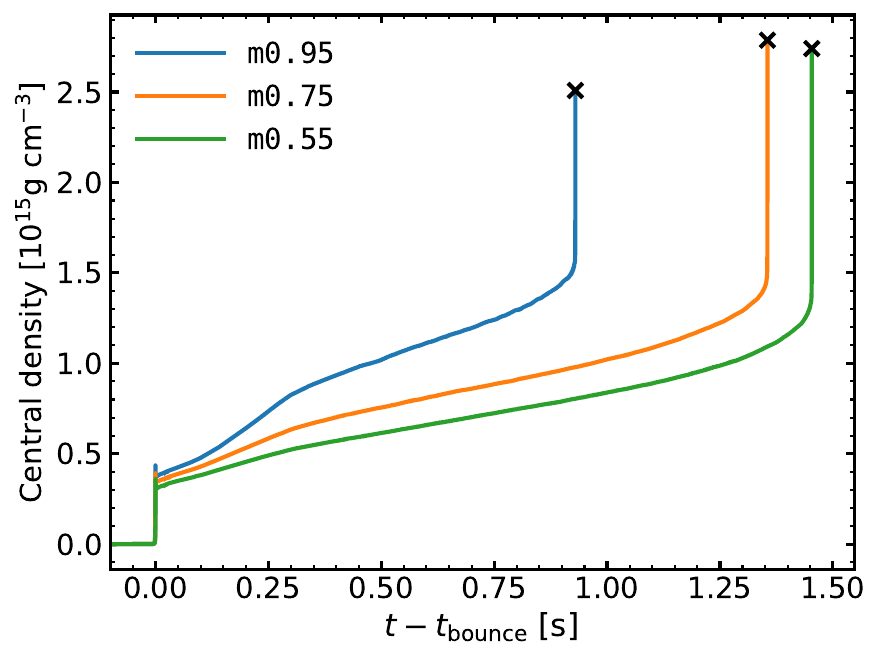}
    \caption{Central density evolution for the three models varying in effective mass. Increasing the effective mass reduces the thermal pressure, leading to a more compact PNS that contracts more rapidly, resulting in an earlier collapse to a BH. BH formation is signaled by a ms-scale collapse of the PNS (black crosses).} 
    \label{fig:central_density}
\end{figure}

\begin{figure}[ht!]
    \centering
    \includegraphics[width=\columnwidth]{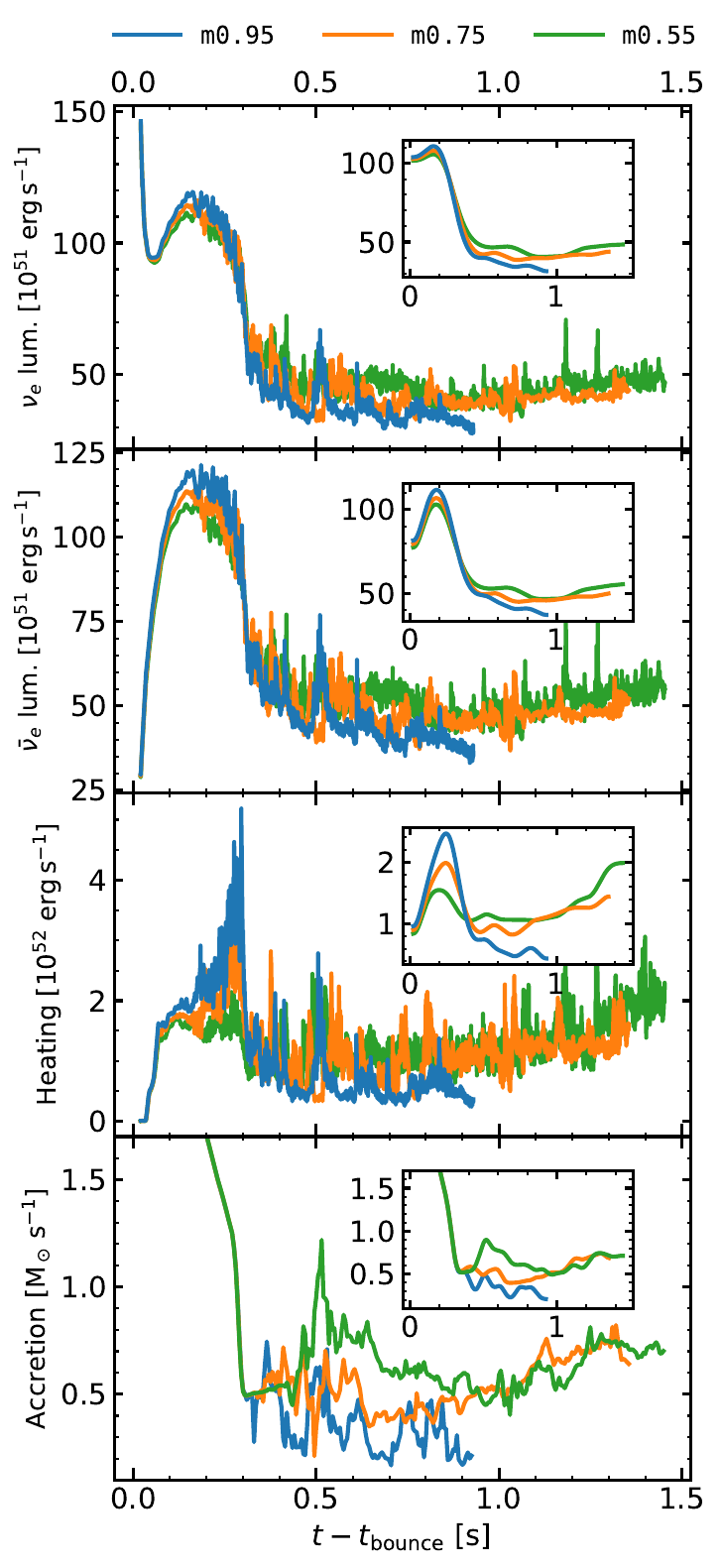}
    \caption{Top to bottom: electron neutrino luminosity, electron antineutrino luminosity, neutrino heating rate, and mass accretion rate. In the figure insets, we show a much more smooth version of the same plots to obtain a clearer view of the trends. With higher effective mass, there is an increase in the displayed neutrino quantities until shock revival, at which point this trend reverses due to the difference in mass ejection by the initial explosion.} 
    \label{fig:neutrino_accretion}
    \vspace{0.4cm}
\end{figure}

\subsection{Prior to BH formation}\label{sec:prior}

We remind the reader that our models in this work, \texttt{m0.95, m0.75, m0.55}, are distinct from one another via the effective mass parameter in the EOS, and the model name reflects the effective mass value for symmetric nuclear matter at nuclear saturation density. Each model undergoes successful shock revival, and
the maximum shock radii pass through 400\,km at $\sim 0.292$\,s (\texttt{m0.95}), $\sim0.317$\,s (\texttt{m0.75}), and $\sim0.346$\,s (\texttt{m0.55}). Shock revival times, as well as other important quantities, are summarized in Table~\ref{table:simulations}. Shock revival does not occur when the silicon-oxygen interface accretes (which is located at $\sim $6.7\,M$_\odot$), but is rather precipitated when the sharp density drop at $\sim 2.4$\,M$_\odot$ accretes which lies deep inside the silicon layer (although some oxygen is present outside this interface; see Figures\,\ref{fig:compactness},\,\ref{fig:progenitor_composition} and Section~\ref{sect:progenitor}.)\\

Figure~\ref{fig:shock_rad} shows the mean shock radius for each model, both before and after BH formation. The time of BH formation is indicated by the vertical dashed lines. We highlight that shock revival occurs even for the most thermally stiff EOS, \texttt{m0.55}, which, based on previous work, is known to be the EOS least prone to explosion \citep{schneider:19, eggenbergerandersen:21}. Lowering the effective mass increases the thermal pressure inside the PNS, such that the PNS forms less compact and contracts less rapidly. This results in a decreased release of gravitational binding energy in the form of neutrinos, and less neutrino heating, making the model less prone to explode. We attribute the successful shock revivals to the high compactness of this 60\,M$_\odot$ progenitor, which has a shallow-density profile. This results in a high accretion rate, which powers a high neutrino luminosity and a hard neutrino spectrum. The hard neutrino spectrum, combined with a high gain region density, results in a high optical depth in the gain region. This high optical depth, along with the high neutrino luminosity, manifest in significant neutrino heating and turbulent convection, enabling shock revival even for the most thermally stiff model.

Figure~\ref{fig:central_density} shows the central density evolution. The higher the effective mass, the more rapid the central density evolves due to a decreased thermal pressure within the PNS. Owing to this thermal pressure dependence, a higher effective mass model can support a less massive PNS and collapses into a BH sooner. Model \texttt{m0.95} forms a BH at $\sim0.93$\,s, model \texttt{m0.75} at $\sim1.36$\,s, and model \texttt{m0.55} at $\sim1.45$\,s.

From top to bottom in Figure~\ref{fig:neutrino_accretion}, we show the electron neutrino luminosity, the electron antineutrino luminosity, the neutrino heating rate, and the mass accretion rate. The neutrino luminosities are the energy flux in the radial direction integrated over a spherical surface at 500\,km. The neutrino heating is the integral over zones, with entropies above 6\,k$_\mathrm{B}$\,baryon$^{-1}$ and densities below $3\times10^{10}$\,g\,cm$^{-3}$, that gain energy (internal plus kinetic) after a time-step in the neutrino unit. The mass accretion rate is the mass per unit time that falls inward through a spherical surface at 500\,km. In the figure insets we show a smoothened version of the same plots to discern trends between the models. We see that the neutrino luminosities and heating rate increase with effective mass before shock revival, such that the \texttt{m0.95} model explodes first and more energetically, and within a window of $\sim50$\,ms, is followed by the \texttt{m0.75} model and then the \texttt{m0.55} model. Note also how this energy hierarchy between the models inverses after shock revival: the higher the effective mass, the more matter is ejected by the initial shock. This decreases the subsequent accretion rate, powering less neutrino emission and heating. It is important to note this energy hierarchy to fully interpret the impact of the thermal EOS, since these effects will have some correlation with the final explosion energy - a topic we revisit at the end of this section.

We observe, but omit to show with a figure, that the higher the effective mass, the higher the average neutrino energy of all species, both before and after shock revival. This occurs because the neutrinospheres are located at higher temperatures \citep{schneider:19}. In contrast to the electron neutrino and electron antineutrino luminosities, the heavy lepton neutrino luminosity increases with the effective mass throughout. This highlights how accretion aids in fueling the emission of electron neutrinos and electron antineutrinos, while the contraction of the PNS fuels the heavy lepton neutrinos.

\begin{figure}[ht!]
    \centering
    \includegraphics[width=\columnwidth]{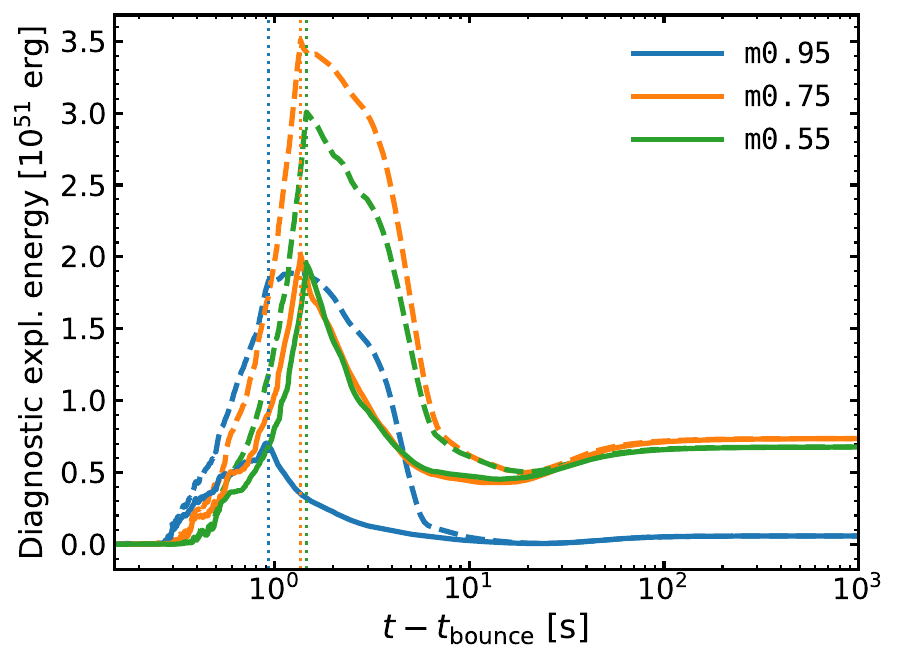}
    \caption{Diagnostic explosion energy according to Equation~\ref{eq:diag_expl} (solid lines). The dashed lines are the diagnostic calculated by replacing the potential energy term, $\Phi$, in Equation~\ref{eq:diag_expl} with  $-GM^{\text{grav}}_\mathrm{encl}/r$ (see text). BH formation occurs at similar times (vertical lines) for the \texttt{m0.75} and \texttt{m0.55} models, resulting in similar explosion energies.  In contrast, the \texttt{m0.95} model experiences BH formation earlier, allowing less time to build up explosion energy via neutrino-driven winds.} 
    \label{fig:explosion_energy}
\end{figure}

\begin{figure}[ht!]
    \centering
    \includegraphics[width=\columnwidth]{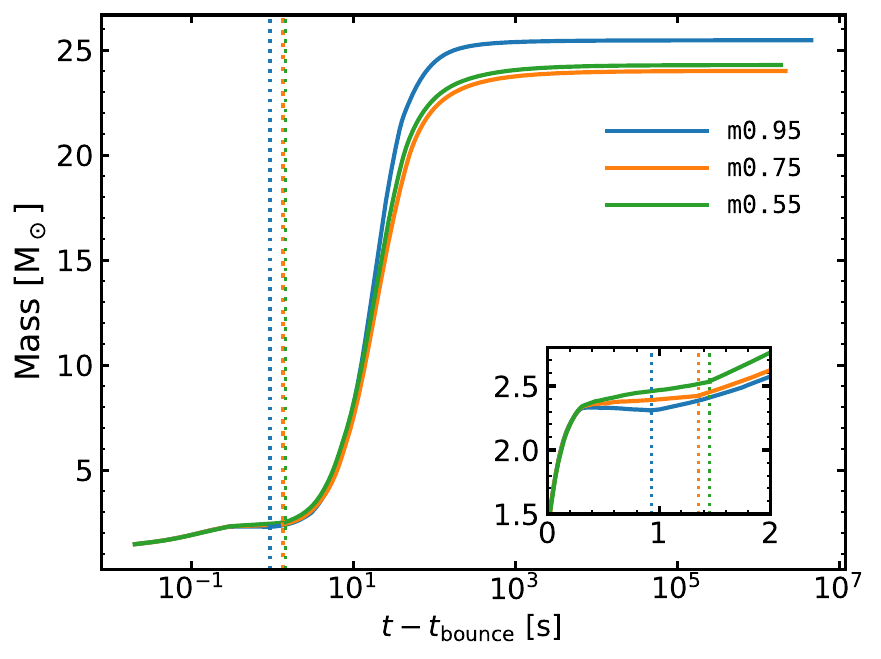}
    \caption{Gravitational mass of the compact object. The vertical lines indicate when the BHs form. The inset zooms in on the first two seconds. After shock revival, higher effective mass models accrete mass more slowly since they eject more material with the revival. $\sim 22\,$M$_\odot$ accretes onto the BHs within hundreds of seconds.} 
    \label{fig:m_grav}
\end{figure}

In Figure~\ref{fig:explosion_energy}, we show the commonly used diagnostic explosion energy, $E_\mathrm{diag}$  \citep{buras:06b,mueller:17}, with the extra condition that the radial velocity is positive (as in \cite{rahman:22}),

\begin{equation} \label{eq:diag_expl}
    E_\mathrm{diag} = \int_{e_\mathrm{tot}, v_\mathrm{r} > 0} \rho e_\mathrm{tot} dV,
\end{equation}

\begin{equation}\label{eq:etot}
    e_\mathrm{tot} = \epsilon_\mathrm{int} + \frac{v^2}{2} + \Phi,
\end{equation}
where $\rho$ is the density, $dV$ a volume element, $\epsilon_\mathrm{int}$ the internal energy, $v$ the fluid velocity, and $\Phi$ the potential energy. The dashed lines in Figure~\ref{fig:explosion_energy} are calculated using $\Phi= -\text{G}M^{\text{grav}}_\text{encl}/r$, where $M^{\text{grav}}_\text{encl}$ is the enclosed gravitational mass at spherical radius $r$. As both variations of the diagnostic explosion energy are used in the literature, we include both. 
When using the potential, the assumption is that every unbound parcel (positive $e_\mathrm{tot}$) needs to ``climb" out of the potential well, taking all the mass shells outside its current radius into account. This results in a form of double-counting inside the unbound region when several mass layers are unbound \citep{mueller:17}. In other words, a parcel is estimated to escape with an energy of $E_\text{diag}$, but its mass still contributes to the potential when another parcel is evaluated in Eq.\,\ref{eq:etot}. Thus, using the potential underestimates the explosion energy. Using $-GM^{\text{grav}}_\text{encl}/r$ does not take into account the mass outside radius $r$, avoiding double-counting but overestimates the explosion energy. Both these definitions assume that unbound regions will not transfer energy to bound regions (negative $e_\mathrm{tot}$) that stay bound. Nor do they take into account the kinetic energy in the bound regions that can contribute to a final explosion energy (e.g. bound material that moves outwards and interacts). As such they are truly only diagnostic energies and long-term simulations are needed to model the redistribution of energy. At late times and large radii, both diagnostics converge onto the final explosion energy.

Before BH formation, the diagnostic explosion energy continues rising as long as the neutrinos inject energy into the matter in the vicinity of the PNS, momentarily unbinding some of that matter. Matter streams back up towards the shock front due to neutrino heating, aided by buoyancy forces since these winds are low in density while high in temperature and entropy. Neutrino-driven winds typically occur along the poles in our simulations, while the bulk accretion occurs in the equatorial plane and powers the neutrino emission. The multi-dimensional dynamics  is crucial in enabling BHSNe events, allowing accretion to occur while the explosion develops. We further discuss dimensionality in Section~\ref{sect:dimensionality}. At BH formation, the neutrino engine ceases to inject energy and the conservative diagnostic explosion energies reach their peak ($\sim0.7\text{--}2.0\times10^{51}$\,erg). The \texttt{m0.95} model peaks at a much lower energy ($\sim0.7\times10^{51}$\,erg) compared to the other two models which are quite similar, with \texttt{m0.75} peaking the highest ($\sim2.0\times10^{51}$\,erg).

Based on 1D simulations, the expectation was that model \texttt{m0.55} would have the highest explosion energy peak, followed by model \texttt{m0.75}, and then model \texttt{m0.95}. A lower effective mass EOS can support a more massive PNS and therefore delay BH formation. This increases the duration between shock revival and BH formation, during which the neutrino engine can build up the explosion energy. In the initial 1D simulations we performed, as well as in the 1D simulations by \citet{schneider:20}, the BH formation times increased much more linearly with decreased effective mass compared to the 2D simulations presented here. One perspective of the three different paths to BH formation was shown by the central density evolution in Figure~\ref{fig:central_density} where the evolutions for model \texttt{m0.55} and model \texttt{m0.75} are more similar than in 1D. Model \texttt{m0.55} forms a BH latest at $1.45$\,s, closely followed by model \texttt{m0.75} at $1.36$\,s, while the PNS in model \texttt{m0.95} collapses at $0.93$\,s. To explain why this occurs in 2D and why the \texttt{m0.75} model peaks at a slightly higher energy than the \texttt{m0.55} model beyond acknowledging stochasticity\footnote{Several phases in CCSNe are stochastic \citep{cardall:15, mueller:20}. For example, the initial shock revival can be more or less energetic due to stochastic and chaotic factors, which affect the subsequent accretion rate.} in SNe simulations, a more in-depth analysis is required. This is presented now.

\begin{figure}[ht!]
    \centering
    \includegraphics[width=\columnwidth]{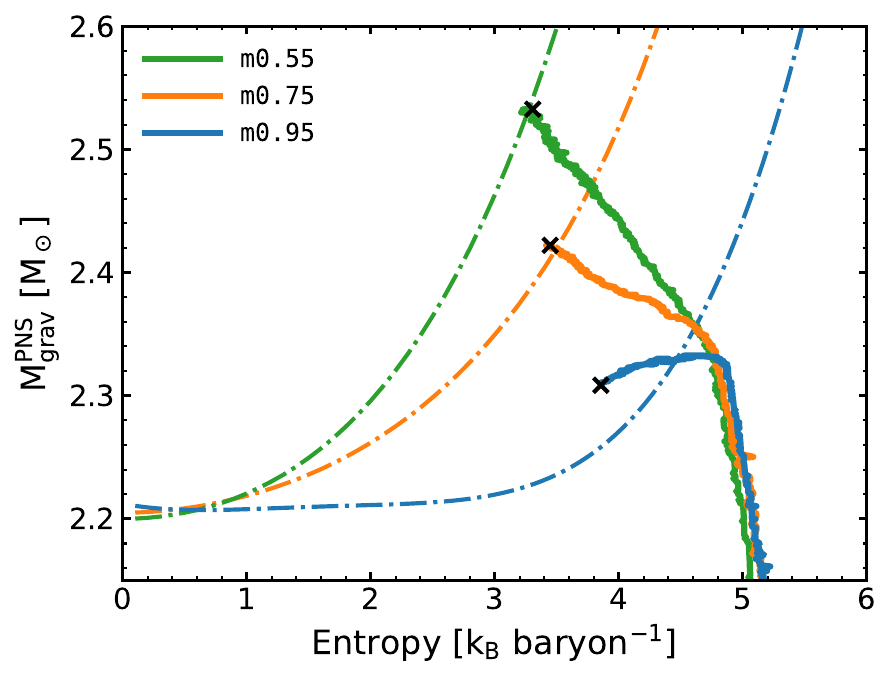}
    \caption{Gravitational mass of the PNS as a function of the most common entropy (solid lines). The dash-dotted lines represent the maximum gravitational mass the EOSs can support given a constant entropy PNS. When the gravitational mass of the PNS reaches the maximum mass set by the EOS (at a certain entropy), collapse is expected to occur \citep{schneider:20}. The black crosses indicate BH formation. After shock revival, the paths in mass-entropy space are broken. The \texttt{m0.75} model has a shallower slope than model \texttt{m0.55} due to a decreased accretion rate, delaying the time to BH formation which impacts the diagnostic explosion energy.}
    \label{fig:mass_entropy}
\end{figure}

\begin{figure*}[ht!]
    \centering
    \gridline{\fig{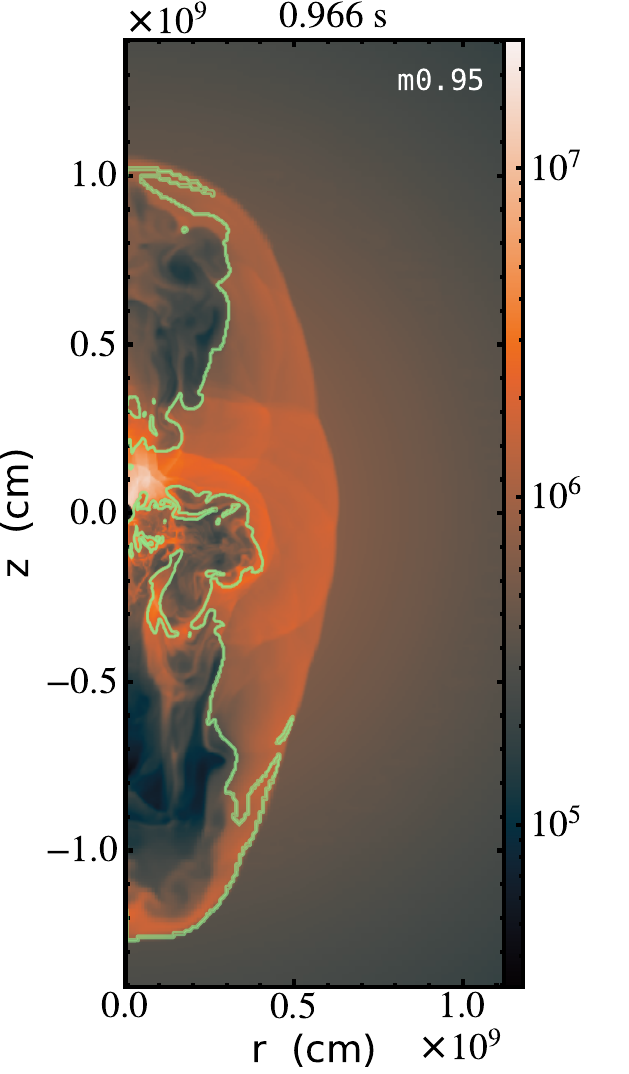}{0.23\textwidth}{}
              \fig{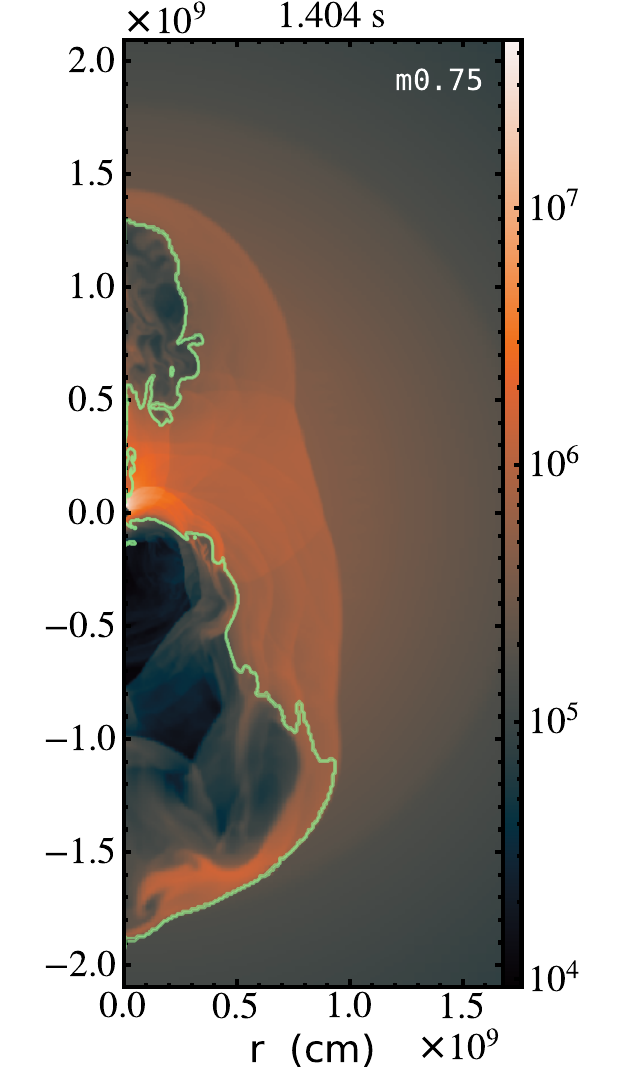}{0.23\textwidth}{}
              \fig{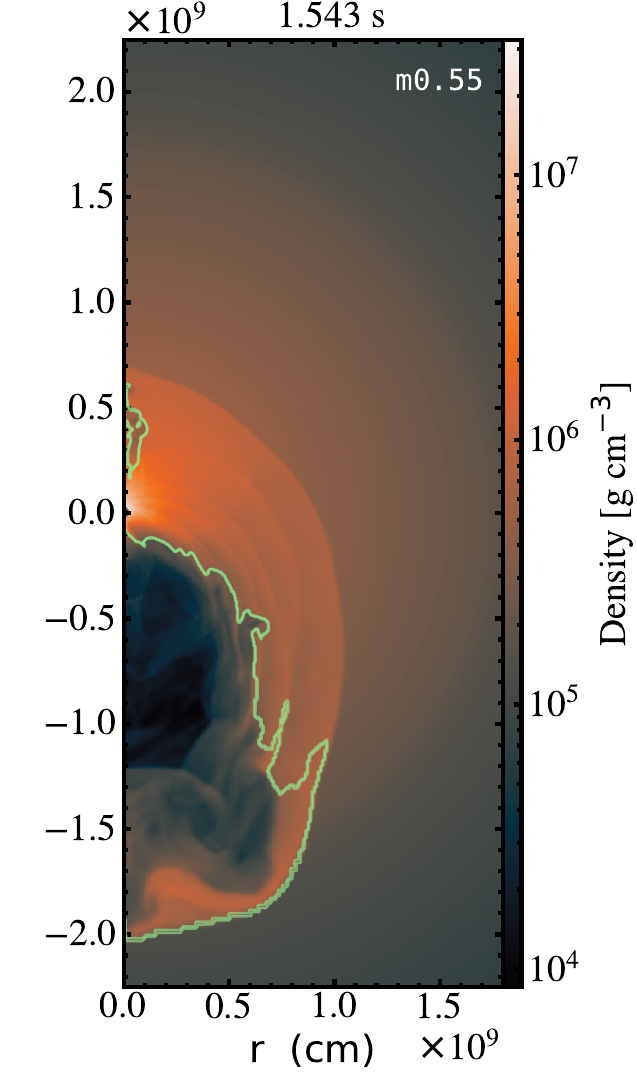}{0.23\textwidth}{}}
    \vspace{-0.5cm}
    \gridline{\fig{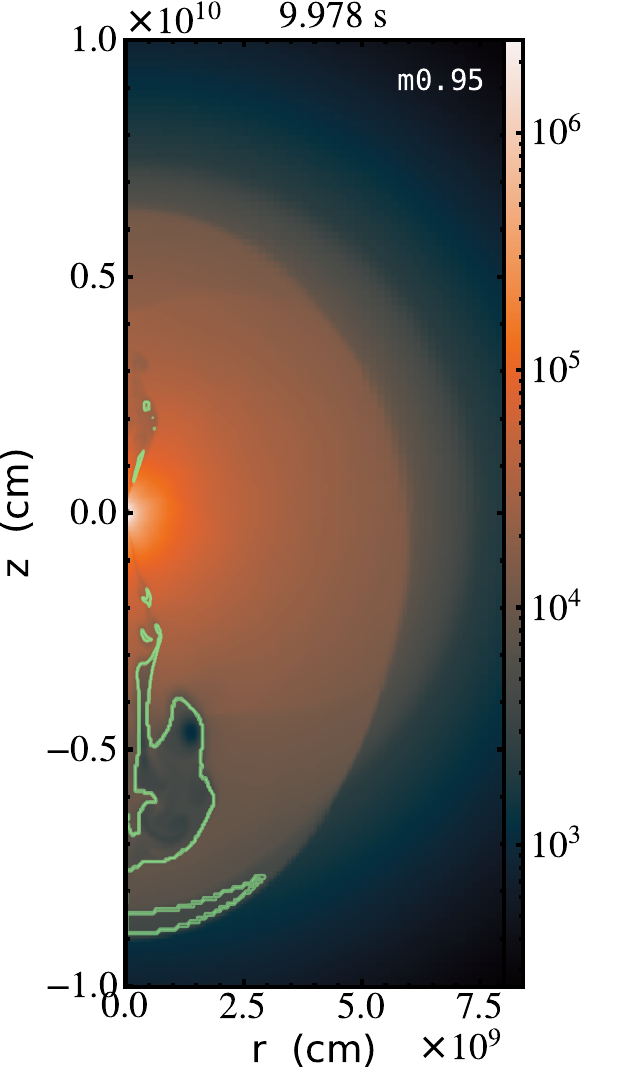}{0.23\textwidth}{}
              \fig{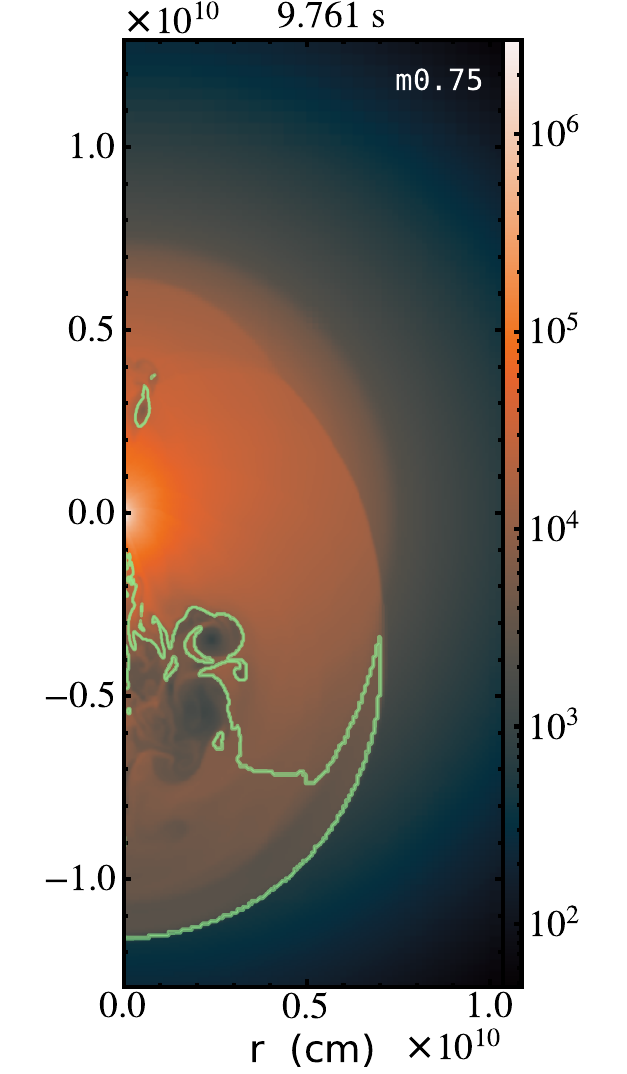}{0.23\textwidth}{}
              \fig{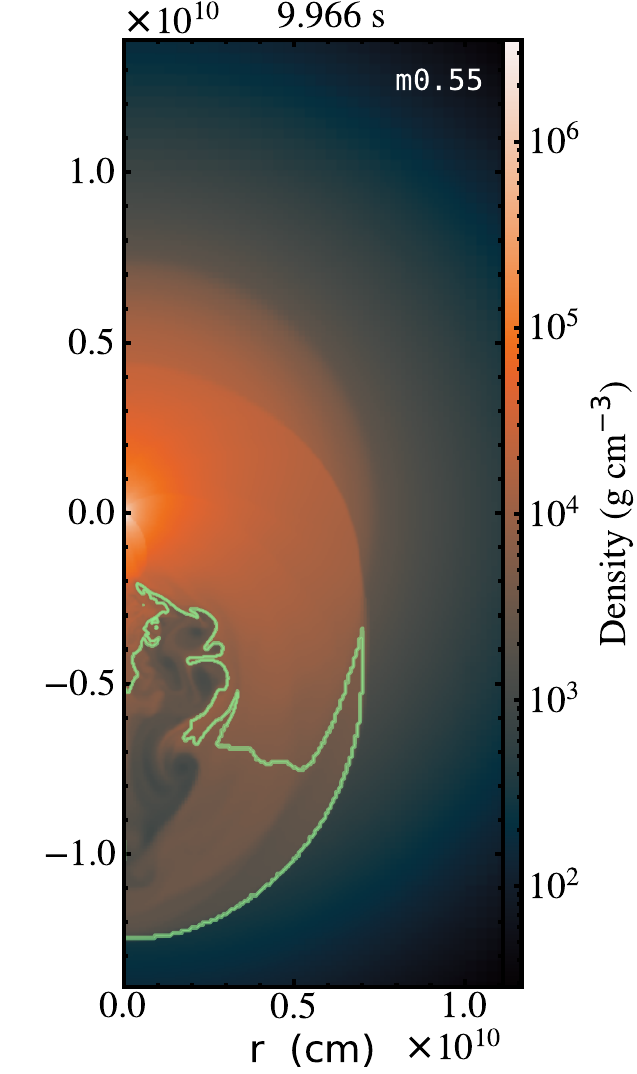}{0.23\textwidth}{}}
    \vspace{-0.5cm}
    \gridline{\fig{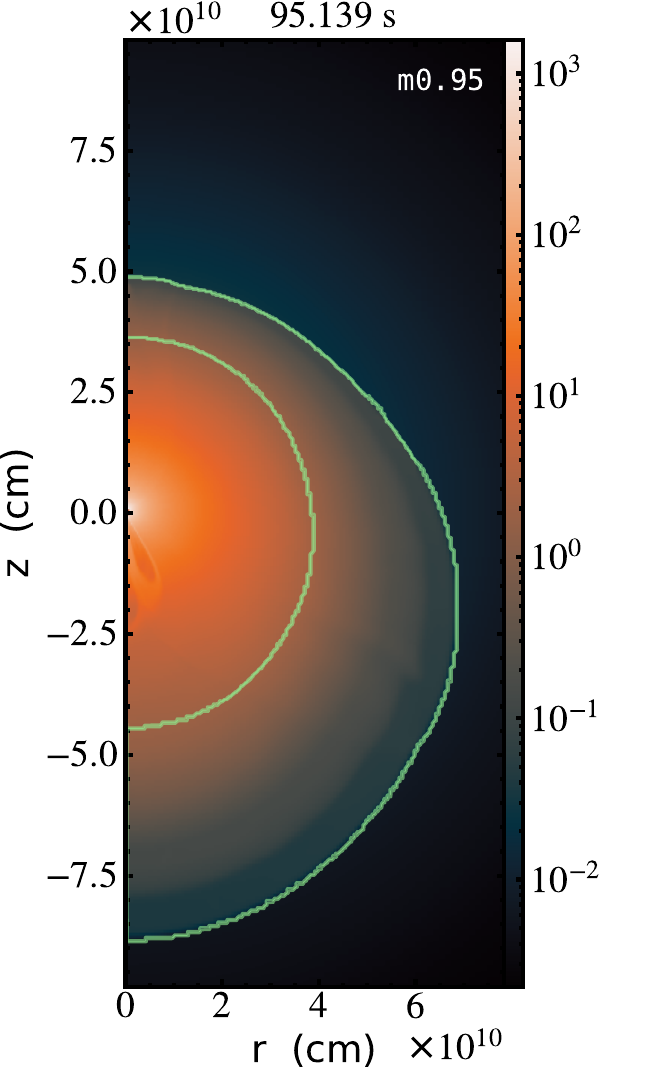}{0.23\textwidth}{}
              \fig{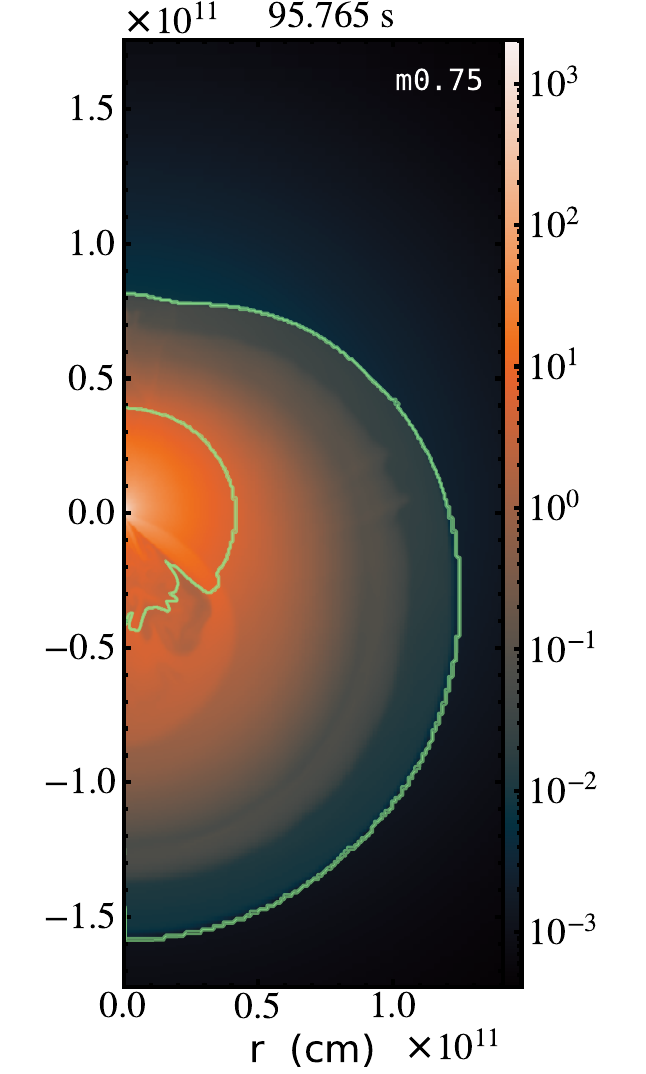}{0.23\textwidth}{}
              \fig{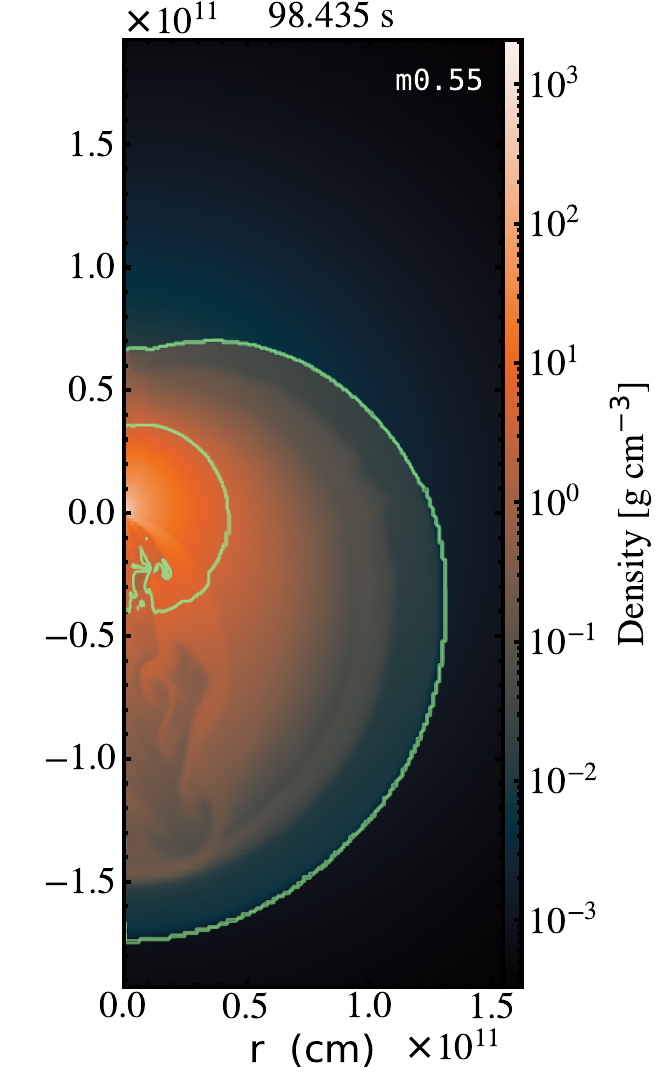}{0.23\textwidth}{}}
    \caption{Density plots showing the explosion morphology and development for each model (\texttt{m0.95}, \texttt{m0.75}, and \texttt{m0.55}, from left to right) at three different stages: BH formation (top), at $\sim10$\,s (middle), and at $\sim100$\,s (top). The green contour separates the bound and unbound regions and demonstrates how the southern unbound region survives and spreads to the northern hemisphere. Note that the scale for the \texttt{m0.95} model is generally smaller than for the other two models.}
    \label{fig:slice_plots}
\end{figure*}

While the explosion energy is most sensitive to the duration between shock revival and BH formation, this interval is set by the compactness evolution of the PNS and is therefore dependent on the accretion rate. Given equal accretion rates, a higher effective mass model has a more rapid PNS compactness evolution. However, after shock revival, the accretion rates differ among the three models. The higher the effective mass, the more matter is initially expelled by the shock and the lower the subsequent accretion rate (recall Figure~\ref{fig:neutrino_accretion}). This impacts the diagnostic explosion energy in two ways which have a tendency to counter each other.
(1) A decreased accretion rate slows down the PNS mass and compactness evolution, delaying the moment when the PNS reaches the maximum mass the EOS can support. Figure~\ref{fig:m_grav} shows the gravitational mass of the compact object evolution over time. In the figure inset, it is clear how the slope of the gravitational mass decreases when increasing the effective mass, increasing the time the neutrino engine is doing work. On its own, this has a positive impact on the explosion energy.
(2) Since accretion fuels the electron neutrino and antineutrino number emission, the heating rate follows the accretion rate which we see in Figure~\ref{fig:neutrino_accretion}. A decreased accretion rate has a negative impact on the slope of the diagnostic explosion energy. The rapid drop in accretion for the \texttt{m0.95} model is likely  causing the shallowing slope in the diagnostic explosion energy after $\sim0.5$\,s (see Figure~\ref{fig:explosion_energy}).

The \texttt{m0.75} model represents the middle case in compactness evolution owing to the thermal part of the EOS. Consequently, \texttt{m0.75} also stands in an intermediate position regarding the amount of matter expelled in the initial explosion, as well as the rates of accretion and heating after explosion. This model achieves a balanced combination by delaying BH formation to build up explosion energy, while maintaining a high enough accretion rate to continue to power the explosion. The result is the highest diagnostic explosion energy peak across the models. Accounting for the small shock revival time difference, the time between shock revival and BH formation only differs by $\sim 61$\,ms between models \texttt{m0.75} and \texttt{m0.55}.

Model \texttt{m0.55}, due to the lower effective mass, requires more mass accumulation to collapse to a BH. Compared to the other models, less matter is ejected during shock revival, resulting in a higher subsequent accretion rate. This decreases the time to BH formation and narrows the gap between models \texttt{m0.55} and \texttt{m0.75}.

We extend the perspective of delaying BH-formation by showing Figure~\ref{fig:mass_entropy}, where we follow \citet{schneider:20} and plot the gravitational mass of the PNS as a function of the most common entropy in the PNS (solid lines). We define the most common entropy as the mass averaged entropy within a window of 2\,k$_\mathrm{B}$\,baryon$^{-1}$ centered on the entropy with the most amount of mass. This is slightly different from the approach in \citet{schneider:20} in order to achieve smooth curves in 2D. The dash-dotted lines in Figure~\ref{fig:mass_entropy} represent the maximum gravitational mass the three different EOSs can support given a constant entropy. \citet{schneider:20} showed that collapse to a BH is expected to occur when these two lines cross, which is what we see. However, compared to the other two models, model \texttt{m0.95} shows a greater deviation between the actual collapse and the line crossing. This deviation is expected for higher effective mass EOSs, as their PNS entropy profiles are more variable and deviate further from being constant \citep{schneider:20}. For all the models, the path in mass-entropy space is broken by shock revival since the accretion rate drops, extending the time to BH formation. Since shock revival occurs closer to the maximum mass set by the EOS for the \texttt{m0.95} model, the accretion rate has less of an impact on the time between shock revival and BH formation, as compared with the lower effective mass models. The shallower slope in the \texttt{m0.75} model, from the lower accretion rate, delays the moment of collapse to a time close to the BH formation time in the \texttt{m0.55} model. The \texttt{m0.55} model has a higher accretion rate and a steeper slope in mass--entropy space, reaching BH formation sooner than it would with a lower accretion rate. \\

\begin{figure*}
\begin{interactive}{animation}{anc/figure10_animation.mp4}
\includegraphics[width=2\columnwidth]{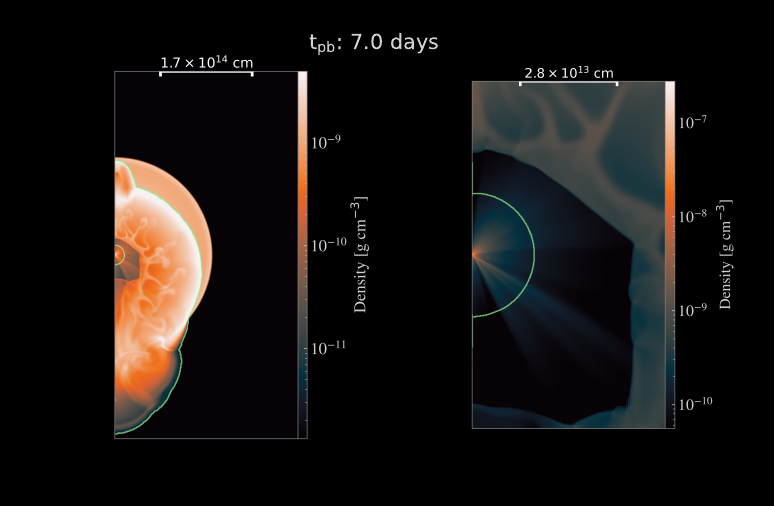}
\end{interactive}
\caption{A snapshot of the density field (at 7 days after core-collapse) from an animation that illustrates the evolution for model \texttt{m0.55} between 20\,ms post-bounce and 22.1\,days. This animation is available as an ancillary file in the \texttt{arXiv} version of the article and can be accessed by clicking on the file from the abstract page. The left panel captures a larger field of view and follows the evolution of the shock, while the right panel emphasizes the area closer to the compact object. The animation illustrates how the standing accretion shock instability \citep{blondin:03} and turbulent convection aid in shock revival around $346$\,ms. Proceeding shock revival, the bulk of the accretion occurs along the equator while neutrino-driven winds are ejected along the poles (particularly in the south), creating an asymmetric explosion morphology. The PNS collapses into a BH at $\sim1.45$\,s. After BH formation, the evolution of the unbound region (the diagnostic ejecta) is shown via the green contours. During the first $\sim12.5$\,s after BH formation, energy is drained from the neutrino-heated bubbles. Although the shock moves outwards in all directions, only the southern unbound region stays unbound until the shock enters a low-density neon/oxygen layer. At this point, the unbound region spreads from south to north (see text for details) and the diagnostic explosion energy increases (see Figure~\ref{fig:explosion_energy}). At $\sim1$\,day, an ingoing reverse shock is clearly visible, accompanied by bullet-like structures resulting from Rayleigh-Taylor instabilities \citep{kifonidis:03}. The shock in the southern region starts penetrating the surface of the star at $\sim3.8$\,days, and at $\sim12$\,days shock breakout is completed along all solid angles. The static version of the animation shows the explosion morphology at $\sim 7$\,days (left panel) and how the ingoing reverse shock is about to merge with itself (right panel).}
\label{movie}
\end{figure*}

\begin{figure}[ht!]
    \centering
    \includegraphics[width=\columnwidth]{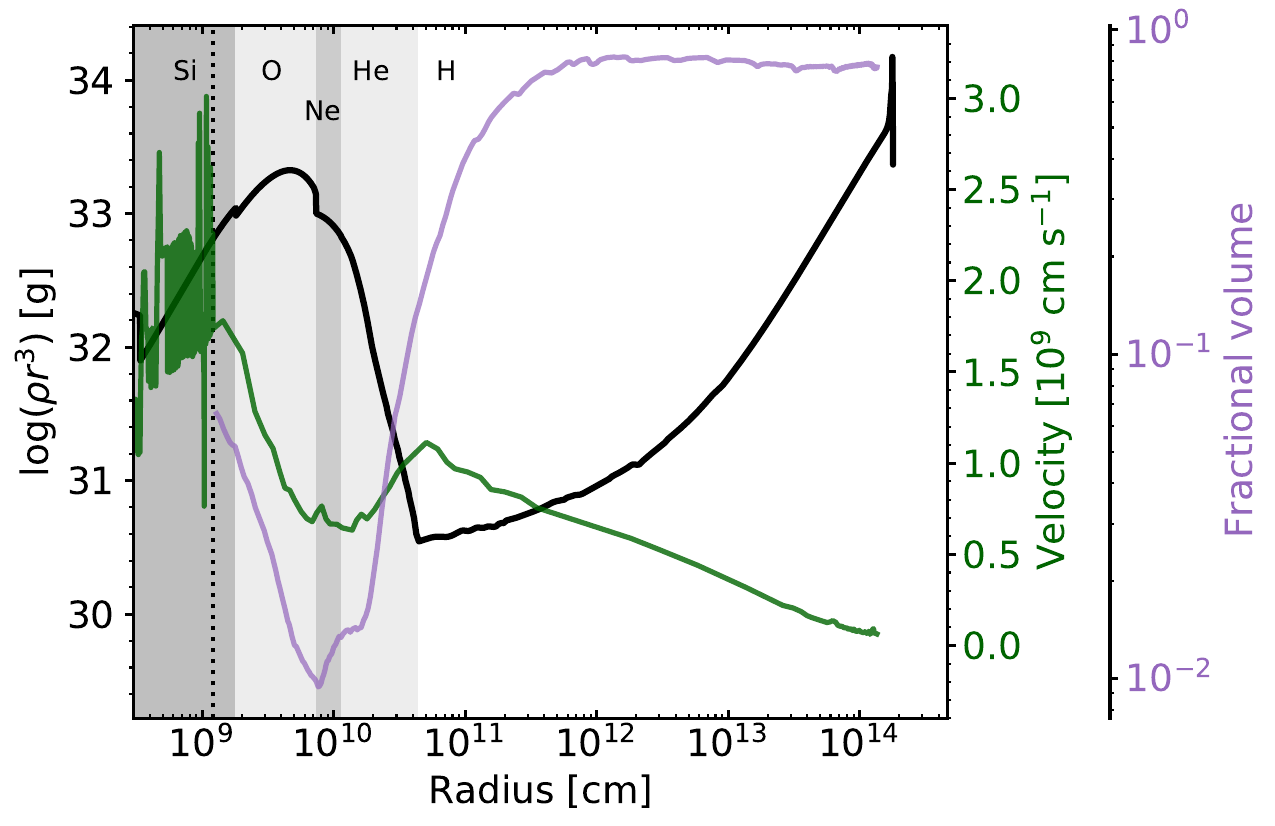}
    \caption{$\rho r^3$ profile of the progenitor (black), the radial velocity of the maximum shock radius (green) as well as the fractional volume of the unbound region (purple) for the \texttt{m0.95} model. The volume is given as a fraction of the spherical volume spanned by the maximum shock radius. Shells are highlighted with shaded bands. The dashed line marks the maximum shock radius at BH formation.} 
    \label{fig:rhor3}
\end{figure}

\begin{figure*}[ht!]
    \centering
    \includegraphics[width=2\columnwidth]{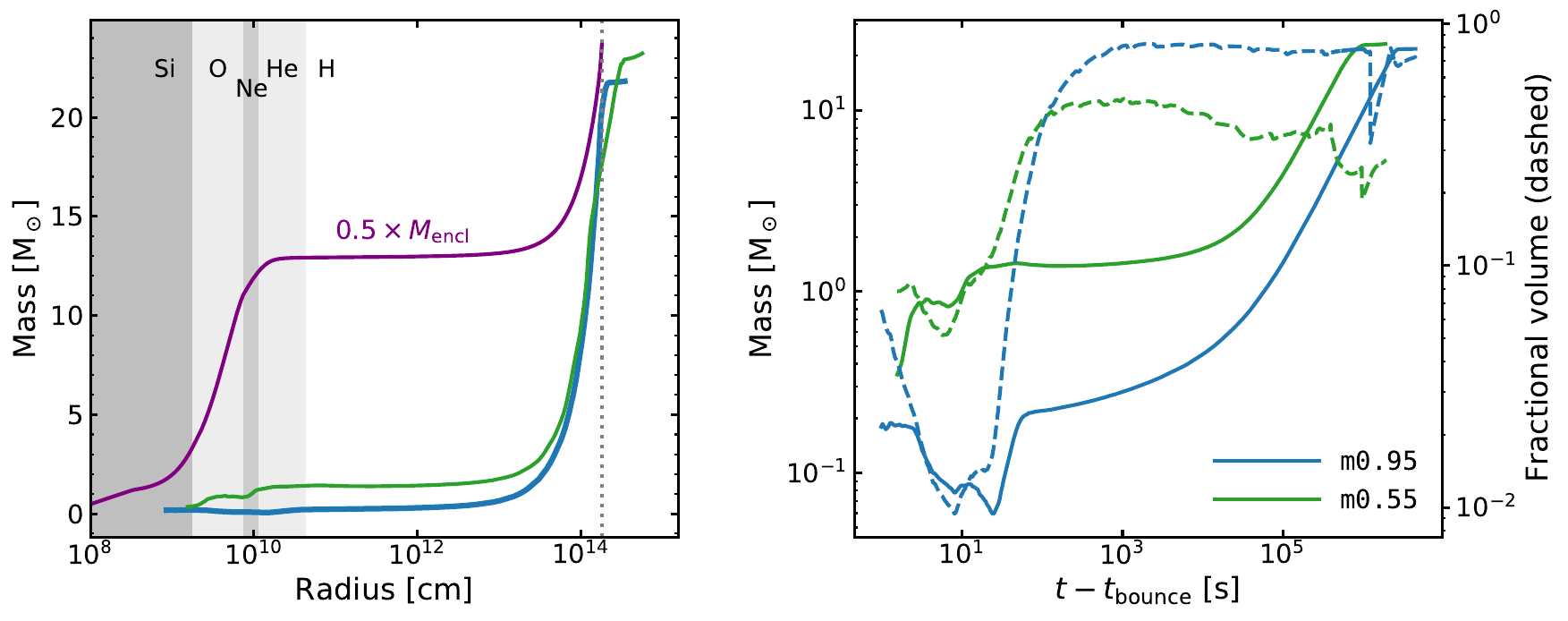}
    \caption{\textit{Left panel}: Enclosed mass (purple; multiplied by 1/2) and the unbound mass for \texttt{m0.95} and \texttt{m0.55} as a function of where the maximum shock radius is. The dotted vertical line marks the edge of the star. \textit{Right panel}: unbound mass as a function of time (solid lines) and the fractional volume of the unbound region (dashed lines; see Figure~\ref{fig:rhor3}). This figure highlights how the unbound mass and volume correlate and are affected by the progenitor structure. During the first $\sim20$\,s, before the unbound regions sphericise above the O--Ne/C interface, the unbound regions suffer mass losses, particularly in the low-energy model \texttt{m0.95}. The unbound ``bubbles" carry $\lesssim1$\,M$_\odot$ from depths where the binding energy is comparatively high -- below the O--Ne/C interface. Above this layer, between $10^{11} - 10^{13}$\,cm, the enclosed mass is nearly constant. All models sweep up the outermost, loosely bound, $\sim21.5$\,M$_\odot$. In model \texttt{m0.55}, due to its higher energy compared to model \texttt{m0.95}, an additional $\sim 1.1$\,M$_\odot$ is brought from low radii before the unbound region spreads along all solid angles and the sweeping phase begins.} 
    \label{fig:unbound_mass}
\end{figure*}

\subsection{After BH formation}\label{sec:afterBH}

Figure~\ref{fig:slice_plots} uses 2D density plots to show the explosion morphology and development for each model (\texttt{m0.95}, \texttt{m0.75}, and \texttt{m0.55}, from left to right) at three different stages: BH formation (top), at $\sim 10$\,s (middle), and at $\sim 100$\,s (bottom). The green contour separates the bound ($e_\mathrm{tot} < 0$ in Equation~\ref{eq:etot})  and unbound regions ($e_\mathrm{tot} > 0$ in Equation~\ref{eq:etot}). It is the unbound regions that contribute to the diagnostic explosion energy (Equation~\ref{eq:diag_expl} and Figure~\ref{fig:explosion_energy}). At BH formation, all three explosions are dipolar, with a preference for the southern pole, especially in models \texttt{m0.75} and \texttt{m0.55}. In 2D, explosions developing along the poles are a common outcome. That the explosion dominates along the southern pole in all three simulations is likely stochastic. Until BH formation, the bulk of accretion occurs along the equator, and, particularly for models \texttt{m0.75} and \texttt{m0.55}, along the northern pole. Meanwhile, as accretion occurs over these solid angles, neutrino-driven outflows emerge along the other solid angles, namely the polar directions, with a stronger southern preference for models \texttt{m0.75} and \texttt{m0.55}, setting the explosion structures at BH formation seen in Figure~\ref{fig:slice_plots}. The unbound regions have high radial velocities, internal energies, entropies, and temperatures, but are low in density. For each model, the total energy in the unbound region at BH formation corresponds to the peak in the model's (conservative) diagnostic explosion energy (see Figure~\ref{fig:explosion_energy}). Since model \texttt{m0.95} has the shortest time between shock revival and BH formation, its unbound regions only have $\sim0.7\times10^{51}$\,erg of total energy, compared to models \texttt{m0.75} and \texttt{m0.55} with $\sim2.0\times10^{51}$\,erg.  After BH formation, the diagnostic explosion energies decline, in part due to these neutrino heated regions doing work and transferring energy to surrounding bound regions which may stay bound and eventually accrete. Furthermore, the diagnostic explosion energies are expected to decline as the shocks sweep up and unbind the gravitationally bound envelopes. Each computational zone in the envelope needs an injection of $|e_\mathrm{tot}|$ amount of energy to become diagnostically unbound. The middle panels in Figure~\ref{fig:slice_plots} show how only the southern unbound region survives at $\sim10$\,s for all three cases, and how the unbound volume is smaller for the low-energy \texttt{m0.95} model. After the decline, in general, the three diagnostic explosion energies reach a minimum between $\sim12\text{--}22$\,s, then rises slightly before asymptoting to a non-zero value after $\sim100$\,s. More specifically, model \texttt{m0.95} reaches its minimum at $\sim22$\,s when the diagnostic explosion energy is $4.28\times 10^{48}$\,erg, having only retained 0.6\,\% of the peak diagnostic explosion energy. Then the diagnostic explosion energy rises and settles at a final explosion energy of $5.7\times 10^{49}$\,erg, which is more than an order of magnitude higher than the minimum. This energy comes primarily from bound matter with positive radial velocity that reaches a layer in the star that is loosely gravitationally bound, unbinding it. We discuss this phenomenon in detail further down in this section and in Section~\ref{sec:binding_energy}. Models \texttt{m0.75} and \texttt{m0.55}, which have higher energy budgets at BH formation and shocks that have reached further out in the ``gravitational well", suffer less fractional energy losses. Model \texttt{m0.75} reaches its minimum at $\sim12$\,s when the diagnostic explosion energy is $4.28\times10^{50}$\,erg, having retained 21.4\,\% of the peak explosion energy. After the rise, its final explosion energy is $7.23\times10^{50}$\,erg. Model \texttt{m0.55} reaches a minimum at $\sim14$\,s when the diagnostic explosion energy is $5.52\times10^{50}$\,erg, having retained 23.6\,\%. Its final explosion energy is $6.68\times10^{50}$\,erg.

The equation of state dependence is also manifested in the ejecta mass and the final BH mass (see Figure~\ref{fig:m_grav}). Model \texttt{m0.95}, which has the smallest time interval between shock revival and BH formation and the lowest explosion energy, ends up accreting more than a solar mass of additional matter onto the BH compared to the other two models. The final BH (ejecta) masses are 25.5 (21.7), 24.0 (23.1), 24.3 (22.8) \,M$_\odot$ for \texttt{m0.95}, \texttt{m0.75}, and \texttt{m0.55}, respectively. See also Table~\ref{table:simulations}.\\

Figure~\ref{movie} shows an animation of nearly the entire simulation for model \texttt{m0.55}. This animation is available as an ancillary file in the \texttt{arXiv} version of the article and can be accessed by clicking on the file from the abstract page. The animation is made from a sequence of 2D density plots and begins at 20\,ms post-bounce and ends at 22.1\,days. The left panel captures the evolution on a larger scale, while in the right panel focus is on the compact object. The animation illustrates how the neutrino-driven winds are ejected along the poles after shock revival, while accretion occurs along the equator. The moment of BH formation is evident at $\sim1.45$\,s by the masking of the central region. After BH formation, the evolution of the unbound region (the diagnostic ejecta) is shown via the green contours. At $\sim1$\,day an ingoing reverse shock is clearly visible, accompanied by bullet-like structures due to Rayleigh-Taylor instabilities \citep{kifonidis:03}. The shock in the southern region starts penetrating the surface of the star at $\sim3.8$\,days, and at $\sim12$\,days shock breakout has occured along all solid angles. The static version of the animation shows the explosion morphology at $\sim 7$\,days (left panel) and how the ingoing reverse shock is about to merge with itself (right panel). \\

The period between $\sim10$\,s and $\sim100$\,s, during which the diagnostic explosion energy increases, represents an interesting phase in the development of the explosion as the shock propagates through the envelope. The middle and bottom panels (at $\sim10$\,s and $\sim100$\,s, respectively) in Figure~\ref{fig:slice_plots} illustrate how, during this time, the unbound regions spread from the southern hemisphere to the northern hemisphere. For model \texttt{m0.55}, the spread of the unbound region is dynamically shown in the online animated version of Figure~\ref{movie}. In general, the rise in each diagnostic explosion energy occurs because bound matter with positive radial velocities can still do work on unbound matter (while becoming ``more bound" itself), as concluded by \cite{chan:18}. Preceding this process, bound matter with positive radial velocity can do work on itself, separating into bound and unbound regions. The onset of these processes are seen around 10\,s in the middle panels of Figure~\ref{fig:slice_plots} -- notice the arrow-shaped features at the leading shock front. Around 10\,s, the leading shock fronts in the southern hemisphere have just entered a low-density layer in the progenitor. Lower density material requires less energy to unbind. The shocks are asymmetric, and as they sequentially enter the low-density layer, the unbound regions spread from south to north. The \texttt{m0.95} model, in particular, for which the diagnostic explosion energy increases by a factor of $\sim10$, demonstrates how the kinetic energy in the outgoing bound matter is sufficient to overcome the binding energy in that layer (and beyond). At $\sim100$\,s the unbound region have sphericised in all models, and the diagnostic explosion energy has settled on its final value.\\

To tie the evolution of both the shock and the unbound region to the progenitor structure, we show the $\rho r^3$ profile as well as the radial velocity of the maximum shock radius as it passes through the progenitor in  Figure~\ref{fig:rhor3} (for the $\texttt{m0.95}$ model). The shock speed responds to the density structure of the progenitor \citep{sedov:59, kifonidis:03}, increasing (decreasing) when the $\rho r^3$ profile decreases (increases). We also plot the unbound volume as a fraction of the spherical volume spanned by the maximum shock radius (as a function of the maximum shock radius). The fractional unbound region size decreases until the maximum shock radius reaches the sharp drop in density at the O--Ne/C interface (at $\sim22$\,M$_\odot$, see Figure~\ref{fig:progenitor_composition}). At that point, the fractional unbound volume reaches its minimum and then starts increasing, spreading across the shock front as the shock front sequentially reaches the interface layer. This is a case of bound matter separating into bound and unbound regions, explaining the rise in the diagnostic explosion energy after $\sim22$\,s. Furthermore, pressure equalization from the high internal energy and high-pressure unbound regions can act much more readily once on the low-density side, which also contributes to the spread of the unbound region. When this happens, we witness a mass flux behind the shock from south to north. Again, this is common in all models.\\

In Figure~\ref{fig:unbound_mass} we highlight the impact of the progenitor structure on the unbound mass for models \texttt{m0.95} and \texttt{m0.55}. The unbound mass can be thought of as the diagnostic ejecta. The right panel shows how the unbound mass as a function of time (solid lines) correlates with the fractional unbound volume evolution (dashed lines), and the impact of the density drop at the O--Ne/C interface on both quantities. In the left panel, the unbound mass is shown as a function of radius (where the maximum shock radius is), and we also include the enclosed mass of the 1D progenitor (purple). When the shock is at $\sim10^{11}$\,cm (around $\sim100$\,s) and the BH masses have nearly asymptoted at $\sim24\text{--}25$\,M$_\odot$ (see Figure~\ref{fig:m_grav}), the explosion energy is carried by only $\sim0.2$\,\,M$_\odot$ and $\sim1.4$\,M$_\odot$ for models \texttt{m0.95} and \texttt{m0.55}, respectively. Then slowly, and once the unbound region has spread along all solid angles, the shock easily sweeps up the loosely bound hydrogen envelope of the star. The final ejecta mass for model \texttt{m0.95} is $\sim21.7$\,M$_\odot$, while model \texttt{m0.55} ejects $\sim22.8$\,M$_\odot$. The $\sim1.1$\,M$_\odot$ difference in the final ejected mass originates from more swept-up matter before the O--Ne/C interface for model \texttt{m0.55}, as well as smaller mass losses before and shortly after crossing that region. The aforementioned mass losses can be seen in the right panel of Figure~\ref{fig:unbound_mass} before $\sim20$\,s. The EOS dependence of the ejected mass would be significantly more pronounced if the hydrogen envelope of this progenitor was smaller.

\subsection{Ejecta composition}\label{sec:ejecta}
\begin{figure*}[ht!]
    \centering
z    \includegraphics[width=0.8\linewidth]{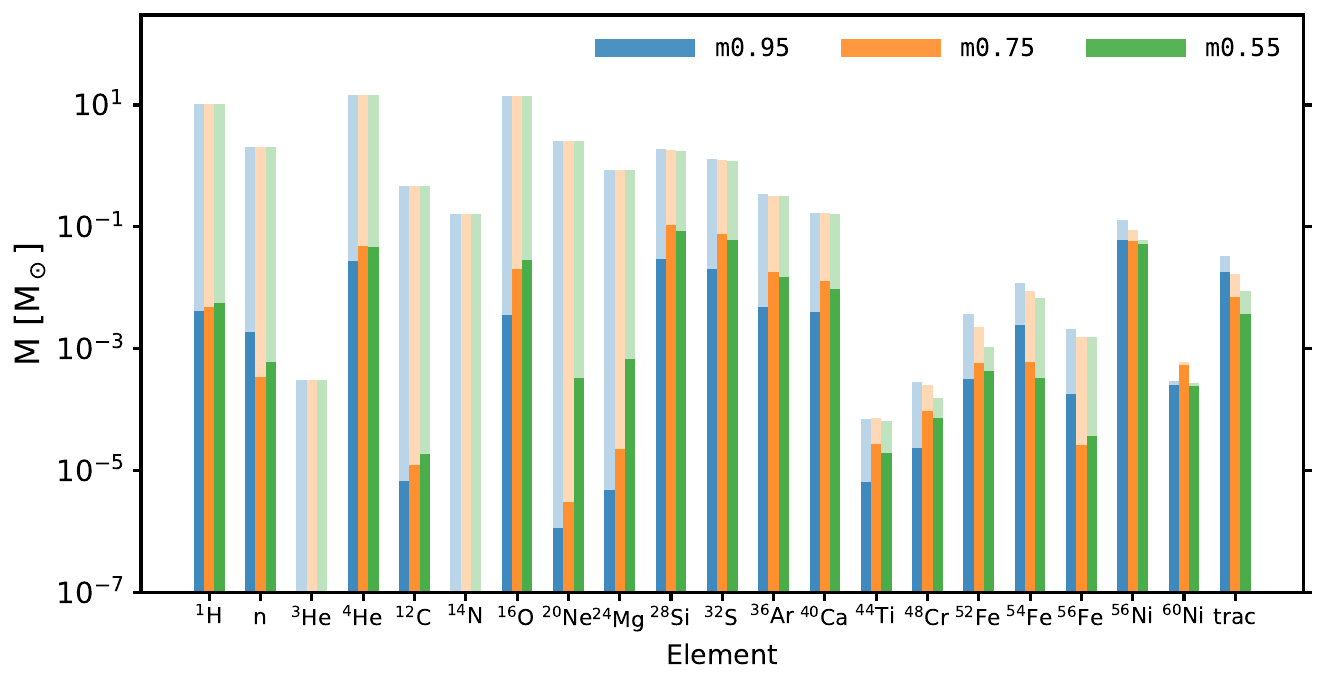}
    \caption{Elemental composition in the unbound region (solid bars) at the time of BH formation. For comparison, we also include the matter on the entire domain (transparent bars). } 
    \label{fig:composition-start}
\end{figure*}

\begin{figure*}[ht!b]
    \centering
    \includegraphics[width=\linewidth]{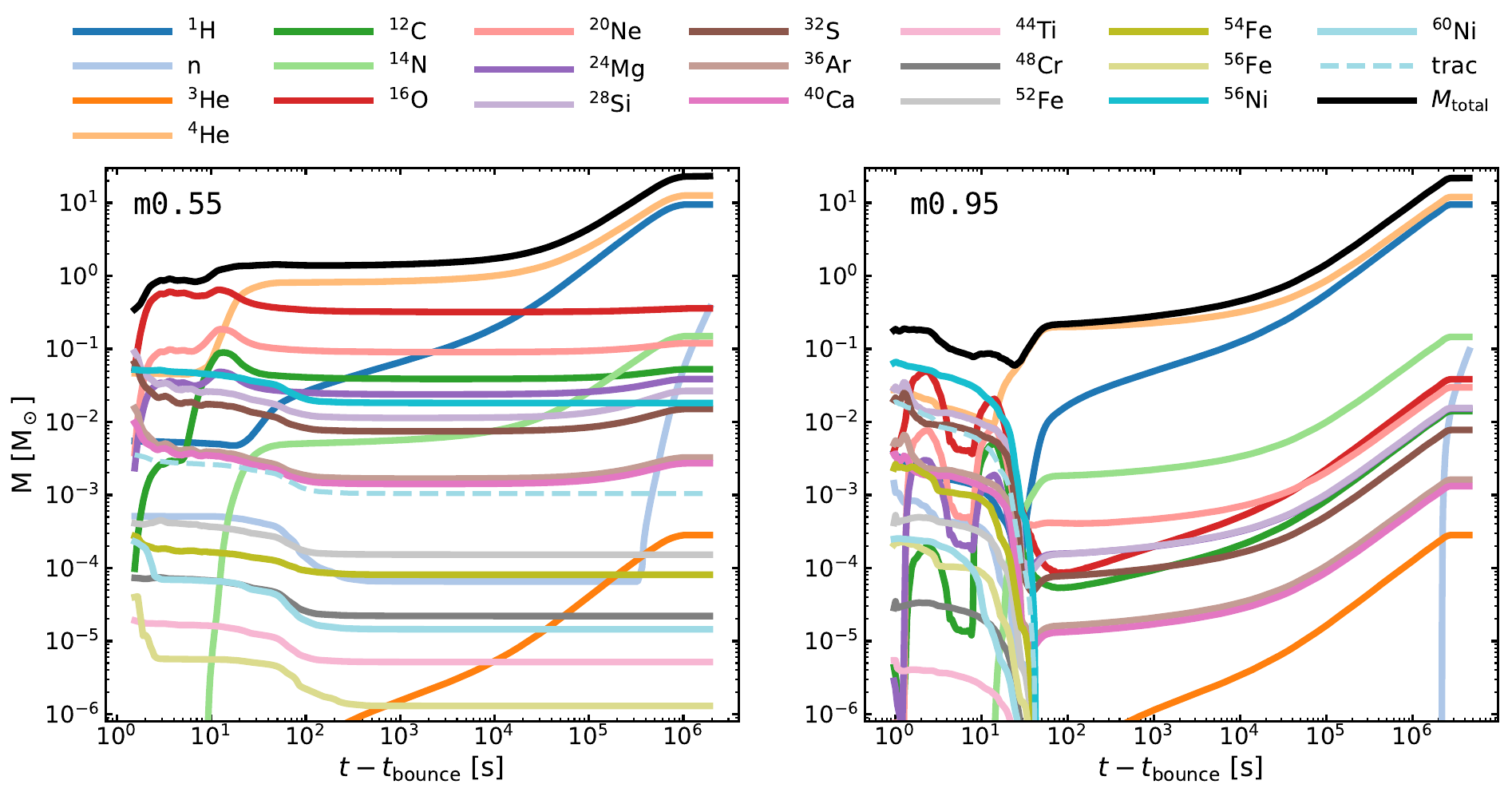}
    \caption{Elemental composition (colored lines) and the total mass (black lines) of the ejecta after BH formation for the \texttt{m0.m5} model (left) and the \texttt{m0.95} model (right). A lot of variability is seen between $1-100$\,s as the high-energy unbound regions interact with bound regions. In this time interval, these evolution curves showcase the interplay between the mass increasing effect of the shock sweeping up material, and the mass decreasing effect of matter leaving the unbound region from below (e.g. the double peaks, see text). For some elements, such as the explosive nucleosynthesis element $^{56}$Ni, there is no additional mass to sweep up as the shock propagates, and a fraction accretes onto the BH. Nearly all the explosively produced elements accrete for the \texttt{m0.95} model, whereas $\sim 0.018\,$\,M$_\odot$ escapes in the \texttt{m0.55} model. The rise of the neutron content around $10^6$\,s is because of the neutron atmosphere we artificially placed outside the star to continue the evolution past shock breakout. See text for further details.} 
    \label{fig:composition-time}
    \vspace{1.3cm}
\end{figure*}

\begin{figure*}[ht!]
    \centering
    \includegraphics[width=0.8\linewidth]{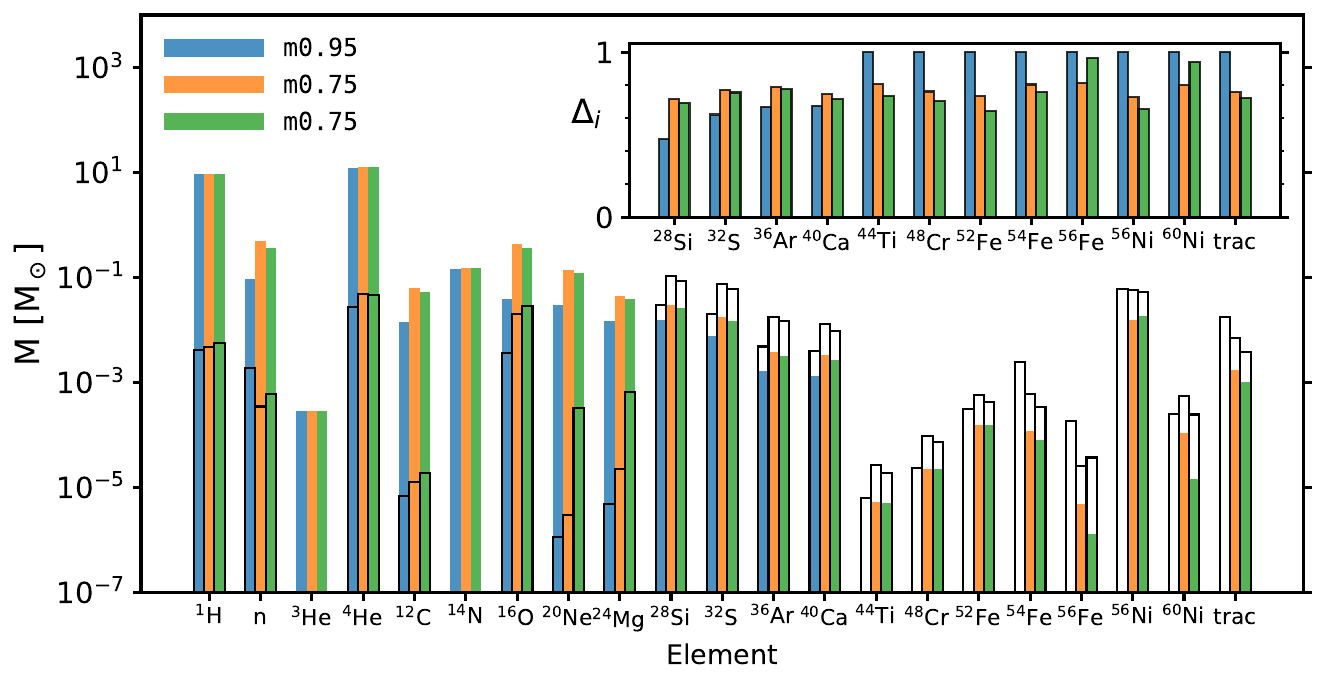}
    \caption{The final composition of the ejecta (solid bars) and the initial composition in the unbound region at BH formation (hollow bars). The inset shows the fraction of mass that is lost (Equation~\ref{eq:fractional_change}) after BH formation. Elements H$^1 - \mathrm{Mg}^{24}$ gain in mass after BH formation, whereas elements Si$^{28} - \mathrm{Ni}^{60}$ decrease in mass. Model \texttt{m0.95} essentially loses all of the $^{56}$Ni that was initially unbound, whereas models \texttt{m0.75} and \texttt{m0.55} lose $73\,\%$ and $66\,\%$, respectively. The EOS dependence of the $^{56}$Ni significantly affects the electromagnetic signal.} 
    \label{fig:composition-end-start}
\end{figure*}

We aim to follow the evolution of the elemental distribution in the unbound material behind the shock front and understand the impact of BH formation on this diagnostic ejecta. Recall that we define a cell to be unbound if the total energy is above 0, or, equivalently, $e_\mathrm{tot} > 0$ in Equation~\ref{eq:etot}. We choose the more conservative measure of $e_\mathrm{tot}$ by using the potential and not $\Phi= -\text{G}M^{\text{grav}}_\text{encl}/r$ in Equation~\ref{eq:etot} (see Section~\ref{sec:prior}). At late times and large shock radii, both measures converge.

In Figure~\ref{fig:composition-start} we show the elemental distribution in the unbound region (solid bars) and of the entire domain (transparent bars; which includes the 2.1\,M$_\odot$ atmosphere tagged as neutrons) for all three models at BH formation and after masking. We show the alpha-chain elements, and combine other species into a tracer element. We have checked that the act of masking has no effect on the $^{56}$Ni mass on the domain. At the respective times of BH formation, the three models, [\texttt{m0.95, m0.75, m0.55}] have 
[0.130, 0.088, 0.061] M$_{\odot}$ of total $^{56}$Ni on the domain, of which [0.061, 0.058, 0.053] M$_{\odot}$ is in the unbound region. In other words, the higher the effective mass, the more $^{56}$Ni has been produced since more mass reached the conditions for NSE in the shock revival phase. There is, however, initially after BH formation, a similar amount of $^{56}$Ni in the models respective unbound regions. To advect this $^{56}$Ni beyond BH formation is a very good approximation since an inconsequential amount of $^{56}$Ni is produced after BH formation. We demonstrate this in Appendix\,\ref{Appendix:A} where, for positions of interest at the time of BH formation, we show the thermodynamic properties along their backward-traced  and forward-extrapolated trajectories on the domain, as well as the evolution of the He$^4$ and $^{56}$Ni abundances.

Figure~\ref{fig:composition-time} shows the evolution of the ejecta composition (the unbound region behind the shock front) from BH formation to 52.3 days for the \texttt{m0.95} model (right panel) and 22.1 days \texttt{m0.55} (left panel) model. At these times, the shock for the \texttt{m0.95} model is still at a lower radius compared to \texttt{m0.55}, but all models are well beyond shock breakout and their ejecta compositions are asymptoted. We omit to show such a plot for \texttt{m0.75}, since this model is qualitatively and quantitatively similar to the \texttt{m0.55} model. These composition curves demonstrate the effect of BH formation on the final ejecta. We see that the fraction of the initially unbound matter that continues to stay unbound after BH formation in this progenitor is mainly dictated in the first $\sim 300$\,s. This can be seen particularly well by looking at the species without any additional mass reservoir in the layers outside the shock at BH formation (like the $^{56}$Ni produced in the explosion), that simply declines and settles at a value that is primarily dictated by the explosion energy at BH formation. Other factors can enter too, such as how far out of the potential well the ejecta has reached when the pressure support from below ceases at BH formation. In general, the closer the shock is to the O--Ne/C interface (see Section~\ref{sec:afterBH} and Figures~\ref{fig:rhor3},\,\ref{fig:unbound_mass}), the lower is the impact of BH formation on the ejecta. For the \texttt{m0.95} model, with the earliest BH formation time and lowest peak explosion energy, many elements undergo a much more significant drop in unbound mass over the initial 100\,s, compared to the other two models. The elements that we impose to have no additional reservoirs in the outer layers (see Section~\ref{sect:methods-nucleosynthesis}), $^{44}\text{Ti--}^{60}$Ni, drop below the plot bounds and essentially reach zero.

The first 100s of seconds also demonstrate how there is one rate of binding matter after BH formation, and one rate of the shock sweeping up new material. To see this, note the double peaked shapes of the \texttt{m0.95} model for e.g. the $^{16}$O. Our interpretation of this phenomenon is as follows. The first peak occurs because the shock and unbound region moves into the $^{16}$O layer that is also enriched by $^{20}$Ne, $^{24}$Mg, and some $^{12}$C (especially further out in this shell; see Figure~\ref{fig:progenitor_composition}). Then there is a general loss of mass and volume in the unbound region. At $\sim 8$\,s, a local minimum in the total mass and for e.g. $^{16}$O occurs, and is concurrent in time with when the maximum shock radius enters the low-density region above the O--Ne/C interface mentioned in Section~\ref{sec:afterBH} and Figure~\ref{fig:rhor3}. Entering this region, when matter is easily ubound, allows for the second peak. Note that the $^{12}$C increases particularly much at the second peak and how this corresponds to the enhanced carbon reservoir in the layer above the O--Ne/C interface at $\sim 10^{10}$\,cm (see Figures\,\ref{fig:progenitor_composition}, \ref{fig:rhor3}). While the unbound region is spreading as it passes through the oxygen poor helium layer between $\sim20\text{--}100$\,s, mass loss from below still occurs. Consequently, the $^{16}$O mass decreases (as does many other elements), while the $^{4}$He content of the ejecta increases. 
When the unbound region has sphericised after $\sim100$\,s, the matter has, to a high degree, separated itself into what is to become ejected and what is not, and the phase of predominantly sweeping up matter follows. During this phase, helium and hydrogen are mostly swept up, constituting the majority of the final ejecta. We see that trace elements corresponding to solar metallicity starting conditions are also swept up. This is particularly evident beyond $\sim100$\,s in the right panel of Figure~\ref{fig:composition-time}, when the mass for several elements heavier than neutrons (due to the artificial atmosphere), hydrogen, and helium slowly increases. Since the helium/hydrogen envelope constitutes the bulk of the ejected mass, we use this mass cut to roughly estimate an ejection of $0.03$\,M$_\odot$ of ``fossil" elements (see Section~\ref{sect:methods-nucleosynthesis}) for each model.

Model \texttt{m0.55} shows similar overall dynamical behaviour for the ejecta composition (left panel of Figure~\ref{fig:composition-time}) as the \texttt{m0.95} model. However, the composition curves are less variable. The higher peak explosion energy, combined with the shock reaching the O--Ne/C interface sooner, results in a lower relative impact on the unbound mass and the individual elements. The net unbound mass loss stops completely once the shock front enters the low-density regime, although elements such as $^{16}$O still become bound from below. While model \texttt{m0.95} looses nearly all its $^{16}$O in the first 100\,s, model \texttt{m0.55} has $\sim0.64$\,M$_\odot$ of unbound $^{16}$O at the peak around 12s and looses about half of it. Similar statements comparing the two models can be made for all other elements except $^{4}$He and $^{1}$H. As such, even with the vast $\sim 20$\,M$_\odot$ of oxygen in this progenitor, only a few tenths of oxygen is ejected for \texttt{m0.55}, and even less for \texttt{m0.95} (essentially only the accumulated solar metallicity elements). Generally, the unbound regions remain in a bubble shape as they pass through the relatively high-density $\sim 20$\,M$_\odot$ oxygen layer. They sphericise only afterwards in the helium layer. Thus, not a large fraction of the available oxygen is swept up. Most of the oxygen core accretes onto the BH instead.\\

Figure~\ref{fig:composition-end-start} shows the final composition of the ejecta after shock breakout for each model (solid bars; i.e. last snapshot in Figure~\ref{fig:composition-time}). We also include the initial unbound ejecta (hollow bars; i.e. first snapshot in Figure~\ref{fig:composition-time}). The figure inset displays the absolute fractional change in unbound mass between these two times,

\begin{equation}
    \Delta_i = |(X^{\mathrm{end}}_i - X^{\mathrm{start}}_i)/X^{\mathrm{start}}_i|,
\end{equation}\label{eq:fractional_change}

\noindent where $X^{\mathrm{start}}_i$ and $X^{\mathrm{end}}_i$ are the initial and final unbound masses for species $i$. Only the elements that lose mass are shown in the inset. In general, the figure shows that elements from $^{1}$H--$^{24}$Mg gain in mass, whereas $^{28}$Si--$^{60}$Ni decrease in mass after BH formation. Model \texttt{m0.95} loses nearly all the mass associated with the heavier species, $^{44}$Ti, $^{48}$Cr, $^{52}$Fe, $^{56}$Fe, $^{56}$Ni, and $^{60}$Ni. In contrast, models \texttt{m0.75} and \texttt{m0.55} retain between $20-35\,\%$ of the initially unbound matter for these species. We comment particularly on the EOS dependence of the $^{56}$Ni, as this affects the electromagnetic signals of these events. Recall that the models [\texttt{m0.95, m0.75, m0.55}] produced [0.130, 0.088, 0.061]\,M$_{\odot}$ of total $^{56}$Ni on the domain, but had a very similar amount in the initial unbound region. The final unbound $^{56}$Ni mass for the three models are [0.0000 ($10^{-17})$, 0.0157, 0.0182]\,M$_{\odot}$, meaning that practically all of the produced $^{56}$Ni in the \texttt{m0.95} model accrete onto the BH, whereas $73\,\%$ and $66\,\%$ of the initially unbound $^{56}$Ni accrete for \texttt{m0.75} and \texttt{m0.55}, respectively.

\section{Discussion}\label{sec:discussion}
\subsection{The explosion energy budget}\label{sec:binding_energy}
\begin{figure}[ht!]
    \centering
    \includegraphics[width=\columnwidth]{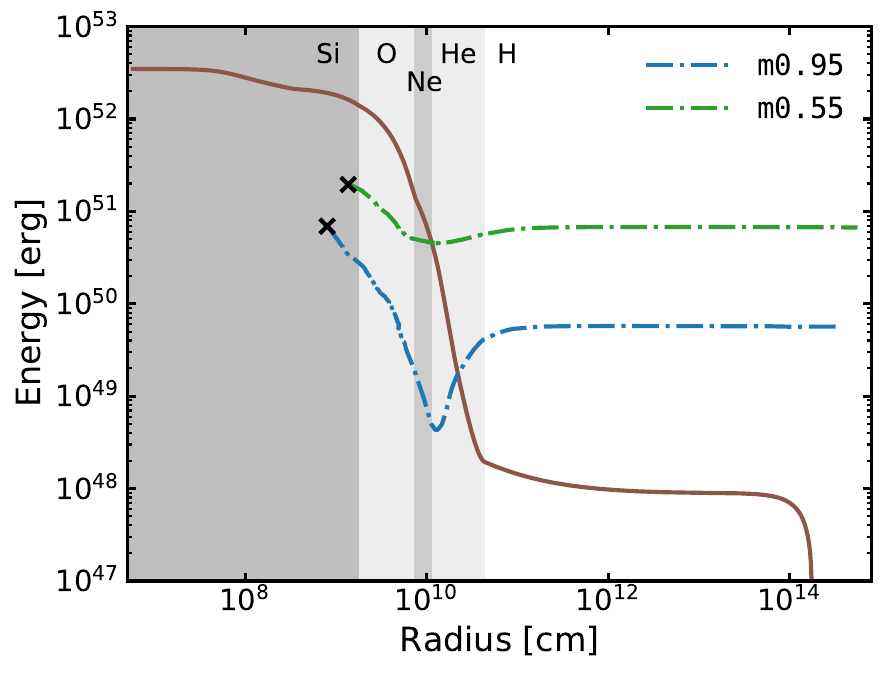}
    \caption{The absolute value of the cumulative binding energy of the progenitor from the edge of the star and inwards (brown line), and the diagnostic explosion energies as a function of where the mean shock radius is for models \texttt{m0.95} (blue line) and \texttt{m0.55} (green line). At BH formation (black crosses), the binding energy outside the shock is more than 10 times the diagnostic explosion energies. }
    \label{fig:binding_energy}
\end{figure}

The core of the 60\,M$_\odot$ progenitor we use is massive and compact. In the simulations presented here, at BH formation, the diagnostic explosion energies are at least a factor of $\sim10$ lower than the binding energy of the material outside the shock. Similar energy imbalances are seen in simulations from \citet{powell:21} and \citet{rahman:22}. Figure~\ref{fig:binding_energy} shows the absolute cumulative binding energy of the progenitor from the surface of the star inward (brown line). The green and blue lines represent the diagnostic explosion energies as a function of the mean shock radius for models \texttt{m0.95} and \texttt{m0.55}, respectively. Only outside the neon layer at $\sim 10^{10}$\,cm, around 10 seconds after shock revival, do the binding energy and the diagnostic explosion energies become comparable. A natural question to pose, then, is how the explosion survives through the massive overburden?

The unbound region in model \texttt{m0.55}  exits the relatively high-binding energy oxygen core with 23.6\% of its initial energy at BH formation. This is a significant amount considering that the binding energy outside the shock at BH formation ($\sim1.62\times10^{52}$\,erg) exceeds by far both the diagnostic explosion energy ($\sim1.95\times10^{51}$\,erg) and the kinetic energy of bound outward-moving matter ($\sim7\times10^{49}$\,erg). We reason that since most of the explosion energy is concentrated in the southern region until $\sim 50$\,s, this bubble of energy does not have to face all of the overburden and transfer all of its energy. Instead, $\sim 22$\,M$_\odot$ accretes onto the BH within $\sim 100$\,s (see Figure~\ref{fig:m_grav}) while there is still only $\sim1$\,M$_\odot$ in the unbound region at this time (see Figure~\ref{fig:unbound_mass}). Once the region with comparably high binding energy is passed, the energy in the unbound region can spread across all solid angles on the other side of the O--Ne/C interface. When this happens, we witness a mass flux from south to north. Only when the unbound region sphericises, and the phase of sweeping up matter begins, does the unbound region ``face" all of the remaining binding energy. 

It is plausible that a shock alone can move outwards in mass coordinate and unbind the loosely bound envelope even with a massive overburden. While the neutrino-heated bubbles in model \texttt{m0.95} aid the propagation of the shock, the rise in the diagnostic explosion energy (after having lost 99.4\%) is due to kinetic energy behind the shock. It is interesting to note that at BH formation, model \texttt{m0.95} has $\sim8\times10^{49}$\,erg of bound matter moving radially outwards, close to the $\sim6\times10^{49}$\,erg rise in the diagnostic when the shock passes the O--Ne/C interface. Model \texttt{m0.95} motivates speculation on scenarios where no parcels with positive energy survive through the overburden \citep[see also][]{rahman:22}. However, once the shock enters into the low-potential energy regime, bound outgoing fluids can give rise to a positive explosion energy by separating itself into ejecta and eventual fallback material, similar to neutrino-induced mass ejection scenarios \citep[e.g.][]{nadezhin:80, fernandez:18}. The \texttt{m0.95} model appears very close to such a case, retaining only 0.6\,\% of the energy in the unbound region after BH formation, until the kinetic energy increases the diagnostic explosion energy by a factor of $\sim10$ (see Section~\ref{sec:afterBH}). Still, model \texttt{m0.95} ejects 21.67\,M$_\odot$. The uncertain outcomes in \citet{powell:21}, \citet{sykes:23}, and to some extent \citet{rahman:22} (see Section~\ref{sec:intro}) are interesting in the light of this discussion. A question that remains is, what dictates whether the shock survives or not? \citet{powell:21} suggested that the shock survives if it has reached the sonic point of the infalling matter at BH formation. The simulations in \citet{sykes:24} are consistent with this hypothesis.

\subsection{2D to 3D}\label{sect:dimensionality}
Supernovae are inherently 3D, and it is important to understand how moving from 2D to 3D affects the results. In 3D, compared to 2D, there is more fine-structure, with smaller but more numerous neutrino heated bubbles and accretion funnels. This geometry can suppress the accretion rate in 3D \citep{mueller:15b, mueller:20, burrows:21}, and therefore act to delay the BH formation time, allowing for the explosion energy to build up longer. However, the lower accretion rate could reduce the slope of the diagnostic explosion energy, competing with the aforementioned effect. Furthermore, 2D models tend to explode more energetically during shock revival \citep{janka:16a}, acting to decrease the subsequent accretion. 3D simulations are required to see which effects dominate over the other within the $\sim 1.5$\,s window we are interested in here. \cite{powell:21} in 3D and \cite{sykes:23} in 2D simulate a high compactness ($\xi_{2.5} = 0.86$), 85\,M$_\odot$ progenitor, where BH formation occurs 100s of ms after shock revival. The times of shock revival and the times of BH
formation are very similar between the two studies, but the peak explosion energy is $\sim 30$\,\% lower in the 2D simulation. We have simulated the same $40$\,M$_\odot$ progenitor in 2D as \citet{burrows:23} use in their 3D simulation. The time of shock revival is similar in our simulation and in the Burrows et al. simulation, but BH formation occurs $200$\,ms earlier in our case. We intend on reporting more details on this simulation, as well as simulations of progenitors spanning a wider mass range, in future work.\\

Another interesting follow-up to this work is to include rotation since $\sim 22$\,M$_\odot$ accrete within the first 100s, after which the accretion rate scales as the predicted $t^{-5/3}$ \citep{chevalier:89b}. This can power outflows and affect the transient via disk formation around the BH. Even for zero net angular momentum models, modelling the stochastic build up of angular momentum can result in a disk \citep{quartert:19}. Here, however, we encounter a limitation of this study due to the axisymmteric assumption where stochastic angular momentum can not be built like in 3D, which yield interesting features in \citet{burrows:23} even with their much lower accretion rate.

\subsection{Final remarks}

It remains an open question of how viable BHSNe are and the mapping relation between progenitors and these outcomes. The mapping is likely mainly a function of the density profile at the time of core-collapse. If these events are realized in Nature for stars that constitute a significant fraction of the distribution of final density profiles, they influence the compact object mass distributions as well as the galactic chemical evolution. Ideally then in the future, BHSNe will also be accounted for in population synthesis calculations.\\

The simulations presented here probe the importance of accretion to fuel the explosion. They hint that even in non-BH forming scenarios, one may end up with a higher final explosion energy if the initial explosion at shock revival is less energetic, given otherwise similar conditions. If the post shock revival accretion rate is decreased by a high-energy explosion, less gravitational binding energy is released as a whole to continue fueling the explosion via winds.

\vspace{0.1cm}
\section{Conclusions}\label{sec:conclusions}
Through three 2D simulations of a high-compactness 60\,M$_\odot$ progenitor, we have explored a recently established BH formation channel, where the BHs have formed between $\sim 0.64$\,s to $\sim 1.11$\,s after shock revival. At these times, the shock waves are still inside the silicon layer, yet all have culminated in successful supernovae. In these explored scenarios, it is the continued accretion of the progenitor that pushes the PNS above its maximum mass, and not fallback material as in fallback supernovae. In contrast to fallback supernovae, there is a large impact on the explosion dynamics. This motivates placing this BH formation channel in a category of its own, and we refer to them as black hole supernovae (BHSNe) in this paper. At BH formation, we excised the central regions while accounting for the accreted mass and followed their evolution for at least 22 days post-collapse. This period extends beyond shock breakout and into a regime where the ejecta composition has settled on its final values.

We have studied the thermal EOS dependence of BHSNe by varying the effective mass of nucleons, following \citet{schneider:19}. We find a strong EOS dependence in the explosion dynamics, the mass of the BH remnant, the ejecta mass and its composition. This highlights the importance of the EOS in these events. Increasing the effective mass, $m^\star$, yields a lower thermal pressure, which results in a faster BH formation time. This is particularly evident when comparing the \texttt{m0.95} model ($m^\star = 0.95$) to the other two. The PNS in model \texttt{m0.95} collapses to a BH at $\sim0.93$\,s after bounce, at which point the diagnostic explosion energy peaks at $\sim0.7\times10^{51}$\,erg. For models \texttt{m0.75} and \texttt{m0.55}, BH formation occurs at 1.36\,s and 1.45\,s, respectively, with similar peak energies around $2\times10^{51}$\,erg. The final explosion energies for models \texttt{m0.95}, \texttt{m0.75}, and \texttt{m0.55} are $\sim 0.06\times10^{51}$\,erg, $\sim 0.72\times10^{51}$\,erg, and $\sim0.67\times10^{51}$\,erg, respectively. We conclude that the final explosion energy and dynamics post-BH formation are primarily determined by the time interval between shock revival and BH formation. Contrary to what one would expect from 1D simulations \citep{schneider:20}, we find that the \texttt{m0.75} and \texttt{m0.55} models peak at similar explosion energies. The \texttt{m0.75} model, being more explosive than the \texttt{m0.55}, expels more matter during the shock revival phase, thereby accreting less and delaying BH formation. Consequently, these two simulations are qualitatively and quantitatively similar. We interpret this similarity with caution since SNe are stochastic due to the turbulent nature of the explosion, exacerbated by the 2D hydrodynamics.\\

For each simulation, the binding energy outside the shock at BH formation exceeds the diagnostic explosion energy by more than an order of magnitude. Therefore, the phase between BH formation and $\sim100$\,s, before the shock has entered the loosely bound envelope at every solid angle, is not trivial. We have paid particular attention to how the unbound region evolves during this critical phase. While $\sim22$\,M$_\odot$ accretes during the first $\sim100$\,s, the mass in the unbound region, located in the southern hemisphere, is $\lesssim1.1$\,M$_\odot$ in each simulation. This multidimensional effect allows the high-energy unbound bubble to circumvent facing all of the overburden, such that the diagnostic explosion energy never drops to zero. In general, when the asymmetrical shock sequentially enters the low-density region as it passes the O--Ne/C interface between $\sim10\text{--}100$\,s, the unbound region spreads to all solid angles, and the diagnostic explosion energy rises. After $\sim100$\,s, matter is separated into ejecta and eventual accretion material, and the phase of slowly sweeping up the loosely bound outer $\sim21.5$\,M$_\odot$ envelope follows. 

The simulations presented here indicate that even if the diagnostic explosion energy drops to zero and no parcel is momentarily unbound, the loosely bound envelope can still be ejected if the shock survives. This is especially evident for the low-energy model, \texttt{m0.95}, which loses 99.4\% of the diagnostic explosion energy after BH formation. Even so, the kinetic energy behind the shock drives up the diagnostic explosion energy again when unbinding the He/H envelope, resulting in a final ejecta mass of $\sim21.7$\,M$_\odot$.

The unbound region for the two higher energy models, \texttt{m0.75} and \texttt{m0.55}, undergo less relative mass and energy losses while passing through the oxygen core where the binding energy is high.
The roughly $1.1$\,M$_\odot$ of material brought from low depths by the unbound bubbles carry the signatures of explosive nucleosynthesis, such as $^{56}$Ni.

We have carefully tracked the composition of the unbound region and contextualized our understanding of the variable evolution curves of the different elements based on insights gained from the unbound region analysis.
At BH formation there is a similar amount of $^{56}$Ni in each model's unbound region, although the total $^{56}$Ni produced increases with the effective mass. Effectively none of the heavy elements, from $^{44}\text{Ti}\text{--}^{60}\mathrm{Ni}$, including $^{56}$Ni, remains unbound after $\sim100$\,s for model \texttt{m0.95}. In contrast, models \texttt{m0.75} and \texttt{m0.55} eject $\sim 0.016$ and $0.018$\,M$_\odot$ of $^{56}$Ni, respectively. These two models, which bring roughly 1\,M$_\odot$ more of material from low radii, leave correspondingly less massive BHs behind.

The different explosion energies, ejecta masses, and compositions (particularly $^{56}$Ni), will give rise to different electromagnetic transients. The duration and quality of the neutrino and gravitational wave signals are also impacted by the thermal EOS. We postpone the study of multimessenger signals from BHSNe to future work. 

If these events are realized in Nature, the EOS affects the mass distribution of stellar-mass BHs through this formation channel. Beyond understanding the impact of the EOS on this formation channel, we encourage more studies on BHSNe in two and three dimensions to unveil its impact on the mass distributions of compact objects. This is an important and timely study, as these distributions are being revealed by data from both gravitational wave interferometers \citep{abbott:23co_distribution} and X-ray observations \citep{oezel:16, lattimer:21}. In the simulations presented here, the masses of the BH remnants are $24.02\text{--}25.48$\,M$_\odot$, rather than the progenitor mass at core-collapse of $\sim$47.3\,M$_\odot$, which is conventionally assumed in a failed SN scenario. Looking ahead, incorporating BHSNe into compact object population synthesis codes \citep[e.g.][]{stevenson:17,giacobbo:18,breivik:20} could prove fruitful.\\

\noindent We thank Matteo Bugli, Daniel Kresse, Finia Jost, Jesper Sollerman, Rodrigo Fernandez, Carl-Johan Haster, and Tuva Marken Ortman for their valuable discussions and insights throughout the course of this work. This work is supported by the Swedish Research Council (Project No. 2020-00452). The computations were enabled by resources provided by the National Academic Infrastructure for Supercomputing in Sweden (NAISS), partially funded by the Swedish Research Council through grant agreement no. 2022-06725. S.M.C. is supported by the U.S. Department of Energy, Office of Science, Office of Nuclear Physics, under award No. DE-SC0017955.

\software{FLASH \citep{fryxell:00}, NuLib \citep{oconnor:15a}, Matplotlib \citep{hunter:07}, NumPy \citep{harris2020array}, SciPy \citep{2020SciPy-NMeth}, yt \citep{turk:11}, h5py \citep{collette_python_hdf5_2014}, SkyNet \citep{lippuner:17b}.}

\appendix

\section{Nucleosynthesis for select positions}\label{Appendix:A}

\begin{figure*}[ht!]
    \centering
    \includegraphics[width=0.51\columnwidth]{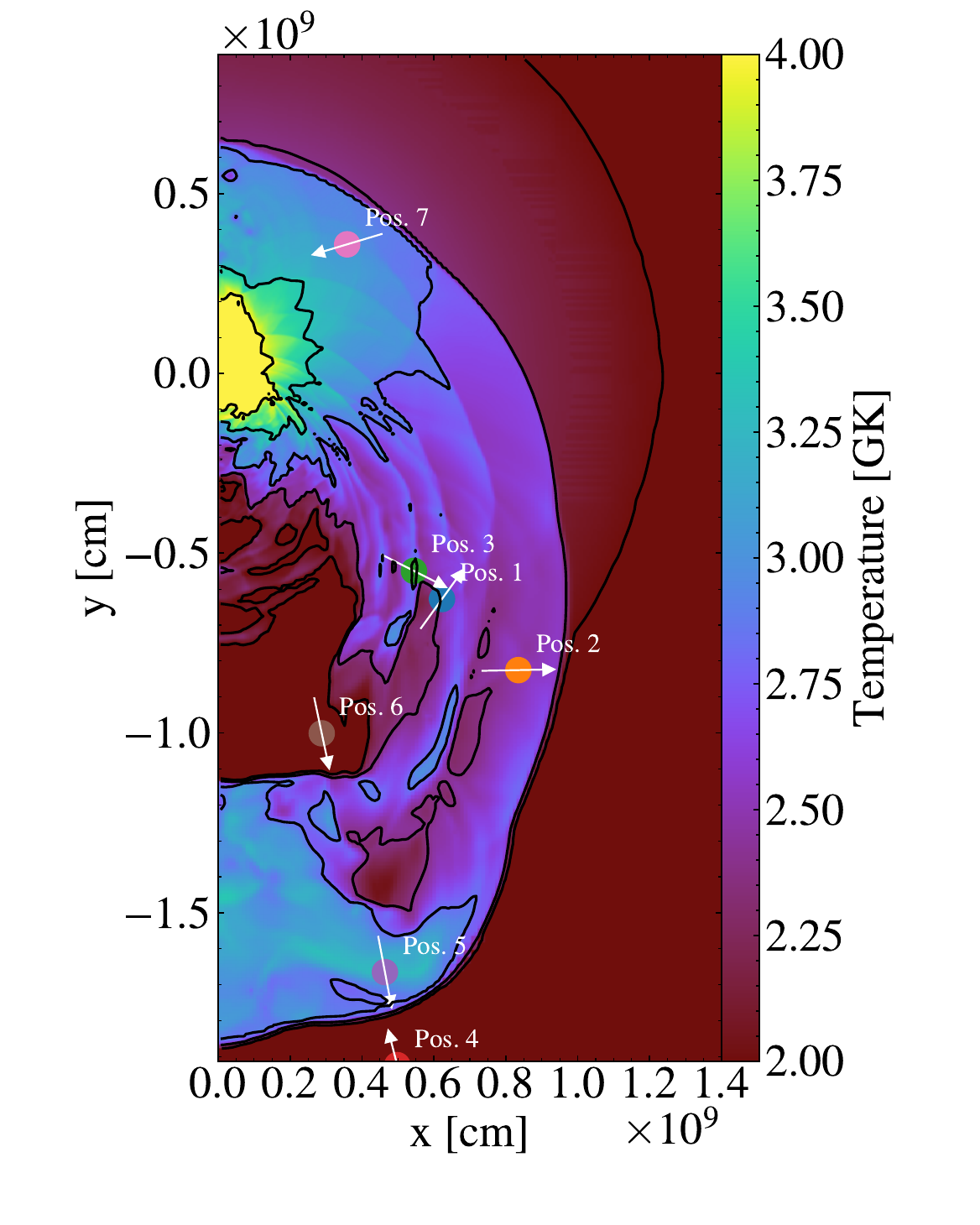}
    \includegraphics[width=0.48\columnwidth]{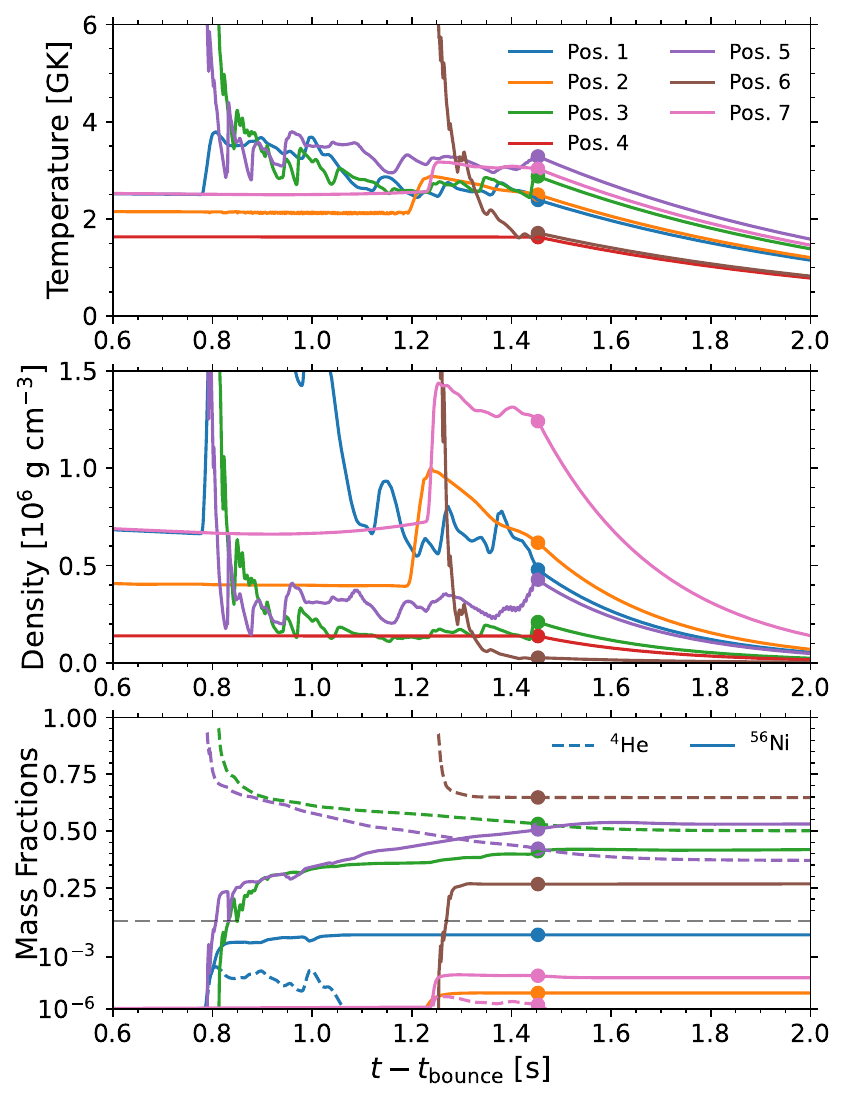}
    \caption{(Left) Temperature color map of the \texttt{m0.55} simulation at the time of BH formation.  The contours mark $T$ = 2, 2.5, 3, 3.5, and 4\,GK, which is also the range of the colour bar.  Seven positions are marked with form the basis for the discussion in this appendix.  (Right) the temperature (top), density (middle), and $^4$He (dashed) and $^{56}$Ni (solid) mass fractions evolutions of each position marks in the left panel are shown. The dots denote the value at BH formation.  The temperature and density following BH formation are set by adiabatic expansion (see text).}
    \label{fig:055_explore_nucleo}
\end{figure*}

As described in \S\ref{sect:methods-nucleosynthesis}, we perform our nucleosynthesis calculation on trajectories obtained from backwards tracing in time starting for the moment of BH formation. In this Appendix, for select positions at the time of BH formation, we show the thermodynamic properties along the trajectory as well as the evolution of the $^4$He and $^{56}$Ni abundance. The select trajectories highlight different regimes of nucleosynthesis and supports our discussion in \S\ref{sec:afterBH} that for the most important unbound ejecta at the time of the BH formation the $^{56}$Ni abundance has reached a steady state.  In the left panel of Figure~\ref{fig:055_explore_nucleo} we show a temperature slice at the time of BH formation for the \texttt{m0.55} simulation.  The contours denote $T=4, 3.5, 3, 2.5,$ and 2\,GK, which is also the range of the colour scale. The shock front at this time coincides with the $2.5\,$GK contour.  We label seven select positions which we explore below. The arrow denotes the direction of the fluid velocity at the time of BH formation. There are several regions. Outside the shock the matter is cold $<2.5\,$GK, position \#4 is in this region. Within the shock, for $y>0$, the fluid velocity is generally negative and matter, while shock heated, is slowly accreting and fuelling the explosion in the south. Position \#7 is in this region. In the interior of the southern region   ($\sim -10^9\,\mathrm{cm} < y<0$ and $x<\sim 0.4\times10^9$\,cm), we have rapidly expanding winds from the PNS. This corresponds to the mass outflow (fueled by accretion elsewhere) that is continually driving the explosion until BH formation.  Due to the rapid expansion, this material has low densities and temperatures.  Position \#6 is in this region. Surrounding this rapid expansion region, but still inside the shock, we have hotter, denser, and more slowly expanding unbound ejecta. Position \#5 is in this region.  
Finally, the $\sim -1.5\times 10^9\,\mathrm{cm} < y<0$, $\sim 0.4\times \sim 10^9<x< \sim 10^9\,\mathrm{cm}$ region is a mixture of expanding shock heated material and ejecta from the PNS. Positions \#1, \#2, and \#3 are in this region.  

In the right panel of Figure~\ref{fig:055_explore_nucleo}, we show the temperature (top), the density (middle), and the $^4$He (dashed) and $^{56}$Ni (solid) mass fractions (bottom) for each of the positions denoted in the left panel and discussed above. We note that, for the composition panel, the scale transitions from linear to logarithmic at a mass fraction value of 0.1. The temperature and density of the trajectory before BH formation (denoted by the point at $\sim$1.45\,s) is determined by our backwards integration.  The temperature and density of the trajectories following BH formation (for the purposes of this appendix) are set by the condition of adiabatic expansion \citep{fowler:64,harris:17},
\begin{eqnarray}
T_i(t) &=& T_{i,\mathrm{BH}} \exp{(-(t-t_\mathrm{BH})/3\tau^*)}\\
\rho_i(t) &=& \rho_{i,\mathrm{BH}} \exp{(-(t-t_\mathrm{BH})/\tau^*)}\,,
\end{eqnarray}
\noindent
where $T_{i,\mathrm{BH}}$ and $\rho_{i,\mathrm{BH}}$ are the temperature and density of the position $i$ at the time of BH formation and $\tau^*$ is taken as 250\,ms for simplicity. This certainly does not capture the hydrodynamic evolution after BH formation correctly, but rather approximately. The composition follows from SkyNet as discussed in \S\ref{sect:methods-nucleosynthesis}. We note that for the main nucleosynthesis results presented in this paper we simply advect the composition determined from SkyNet at the time of BH formation, and therefore we do not utilize such an extrapolation within our simulations. As discussed below, this is reasonable for the unbound ejecta at the time of BH formation.

First we note that for all the trajectories from all the regions, there is little to no $^{56}$Ni production after BH formation.  The small exception are for Positions \#3 and \#5.  These are relatively (compared to Position \#6) slow trajectories of ejecta from the PNS.  At the time of BH formation the temperatures are $\sim3-3.5$\,GK and therefore some $^{56}$Ni production, via $^4$He capture on $\alpha$ elements, occurs within the adiabatic expansion extrapolation regime in these tests.  The remaining positions are too cold (or do not contain enough $^4$He) at this time to effectively produce $^{56}$Ni. Another interesting trajectory is Position \#1.  This trajectory undergoes incomplete burning after reaching peak temperatures of only $\sim$4\,GK at 0.8\,s after bounce.  The nucleosynthesis occurring at this point can be summarized as the remaining $^{16}$O ($\sim$8\%) being explosively burned into $^{40}$Ca, $^{32}$S, and $^{56}$Ni. Ultimately, very little $^{56}$Ni is produced (with a final mass fraction of $\sim$0.017). The majority of the mass at this position is silicon and sulfur (see Figure~\ref{fig:progenitor_composition} at a radius of $\sim 8\times10^8$\,cm, which is the starting radius of this particular trajectory), those mass fractions change very little during the burning of the oxygen. We note that the trajectory does not include the energy generation from the burning itself.

\end{document}